[ english version / português abaixo ]

# Flat-earth communities on Brazilian Telegram:

## when faith is used to question the existence of gravity as a physics phenomenon

*Ergon Cugler de Moraes Silva*

Brazilian Institute of Information in
Science and Technology (IBICT)
Brasília, Federal District, Brazil

contato@ergoncugler.com
www.ergoncugler.com

**Abstract**

Conspiracy theories related to flat-earthism have gained traction on Brazilian Telegram, especially in times of global crisis, such as the COVID-19 pandemic, when distrust in scientific and governmental institutions has intensified. Therefore, this study aims to address the research question: **how are Brazilian conspiracy theory communities on flat earth topics characterized and articulated on Telegram?** It is worth noting that this study is part of a series of seven studies whose main objective is to understand and characterize Brazilian conspiracy theory communities on Telegram. This series of seven studies is openly and originally available on arXiv at Cornell University, applying a mirrored method across the seven studies, changing only the thematic object of analysis and providing investigation replicability, including with proprietary and authored codes, adding to the culture of free and open-source software. Regarding the main findings of this study, the following were observed: During the Pandemic, flat-earthist discussions increased by 400%, driven by distrust in scientific institutions; Flat-Earther communities act as portals for other conspiracy theories, such as the New World Order; Although smaller, the flat-Earther network has influential groups that disseminate content and perpetuate narratives; Religious themes such as "God" and "the Bible" are central, combining religious elements with distrust in science; Flat-Earther communities use themes such as gravity to challenge established scientific concepts, reinforcing an alternative view of the world.

### Key findings

- ➔ During the peak of the COVID-19 pandemic, discussions in Brazilian flat-earth communities on Telegram increased by 400% between 2019 and 2020, driven by distrust in scientific and governmental institutions, reflecting how global crises intensify the spread of disinformation;

- ➔ Flat-earth communities not only attract new followers but also serve as gateways to other conspiracy theories, such as the New World Order and Occultism, with 1,934 links originating from NWO and 797 from Apocalypse and Survivalism, demonstrating the interconnectedness of these narratives;

- ➔ The flat-earth network is relatively smaller compared to other conspiracy theory communities, but a few groups on Telegram act as primary hubs for content dissemination, possessing a disproportionate capacity to influence discourse and perpetuate narratives, while smaller groups orbit around them;

- ➔ Discussions about the New World Order are predominant in flat-earth communities, with 724 links directed towards this theme, highlighting that the narrative of global control is used to legitimize and reinforce flat-earth beliefs, creating thematic overlap;
- ➔ The spread of flat-earth beliefs relies heavily on digital platforms, with terms like "channel" and "Telegram" being central in the discussions. The prevalence of these platforms, particularly Telegram, underscores their importance in the maintenance and expansion of communities;
- ➔ Analyses reveal a strong presence of religious terms like "God" and "Bible" in flat-earth discussions, suggesting that these beliefs are shaped by a worldview that combines religious elements with distrust in science, sustaining alternative narratives;
- ➔ Flat-earth communities frequently use themes such as gravity and the evolution of species to reject established scientific concepts, challenging not only the shape of the Earth but the entire structure of scientific knowledge, reinforcing a group worldview;
- ➔ Themes like the biblical apocalypse and magnetic pole reversal are closely connected to flat-earth beliefs, suggesting that these communities associate the belief in a flat Earth with catastrophic predictions, reinforcing apocalyptic narratives that offer alternative explanations for global crises;
- ➔ Despite efforts to dismantle these beliefs, flat-earth narratives show resilience, maintaining active and cohesive discussions over the years, with high levels of interest indicating a deep-rootedness of these beliefs in the digital environment;
- ➔ The interconnectedness between flat-earth beliefs and other conspiracy theories suggests that engagement with this narrative serves as a test of ideological commitment, exposing individuals to a broader range of conspiracies and complicating the deconstruction of these already entrenched beliefs as a sense of community among its members.

## 1. Introduction

After analyzing thousands of Brazilian conspiracy theory communities on Telegram and extracting tens of millions of content pieces from these communities, created and/or shared by millions of users, this study aims to compose a series of seven studies that address the phenomenon of conspiracy theories on Telegram, focusing on Brazil as a case study. Through the identification approaches implemented, it was possible to reach a total of 30 Brazilian conspiracy theory communities on Telegram on flat earth topics, summing up 632,901 content pieces published between June 2019 (initial publications) and August 2024 (date of this study), with 38,520 users aggregated from within these communities. Thus, this study aims to understand and characterize the communities focused on flat earth present in this Brazilian network of conspiracy theories identified on Telegram.

To this end, a mirrored method will be applied across all seven studies, changing only the thematic object of analysis and providing investigation replicability. In this way, we will adopt technical approaches to observe the connections, temporal series, content, and overlaps of themes within the conspiracy theory communities. In addition to this study, the other six are openly and originally available on arXiv at Cornell University. This series paid particular attention to ensuring data integrity and respecting user privacy, as provided by Brazilian legislation (Law No. 13,709/2018 / Brazilian law from 2018).

Therefore, the question arises: **how are Brazilian conspiracy theory communities on flat earth topics characterized and articulated on Telegram?**

## 2. Materials and methods

The methodology of this study is organized into three subsections: **2.1. Data extraction**, which describes the process and tools used to collect information from Telegram communities; **2.2. Data processing**, which discusses the criteria and methods applied to classify and anonymize the collected data; and **2.3. Approaches to data analysis**, which details the techniques used to investigate the connections, temporal series, content, and thematic overlaps within conspiracy theory communities.

### 2.1. Data extraction

This project began in February 2023 with the publication of the first version of TelegramScrap (Silva, 2023), a proprietary, free, and open-source tool that utilizes Telegram's Application Programming Interface (API) by Telethon library and organizes data extraction cycles from groups and open channels on Telegram. Over the months, the database was expanded and refined using four approaches to identifying conspiracy theory communities:

**(i) Use of keywords:** at the project's outset, keywords were listed for direct identification in the search engine of Brazilian groups and channels on Telegram, such as "apocalypse", "survivalism", "climate change", "flat earth", "conspiracy theory", "globalism", "new world order", "occultism", "esotericism", "alternative cures", "qAnon" "reptilians", "revisionism", "aliens", among others. This initial approach provided some communities whose titles and/or descriptions of groups and channels explicitly contained terms related to conspiracy theories. However, over time, it was possible to identify many other communities that the listed keywords did not encompass, some of which deliberately used altered characters to make it difficult for those searching for them on the network.

**(ii) Telegram channel recommendation mechanism:** over time, it was identified that Telegram channels (except groups) have a recommendation tab called "similar channels", where Telegram itself suggests ten channels that have some similarity with the channel being observed. Through this recommendation mechanism, it was possible to find more Brazilian conspiracy theory communities, with these being recommended by the platform itself.

**(iii) Snowball approach for invitation identification:** after some initial communities were accumulated for extraction, a proprietary algorithm was developed to identify URLs containing "t.me/", the prefix for any invitation to Telegram groups and channels. Accumulating a frequency of hundreds of thousands of links that met this criterion, the unique addresses were listed, and their repetitions counted. In this way, it was possible to investigate new Brazilian groups and channels mentioned in the messages of those already investigated, expanding the network. This process was repeated periodically to identify new communities aligned with conspiracy theory themes on Telegram.

**(iv) Expansion to tweets published on X mentioning invitations:** to further diversify the sources of Brazilian conspiracy theory communities on Telegram, a proprietary search query was developed to identify conspiracy theory-themed keywords using tweets published on X (formerly Twitter) that, in addition to containing one of the keywords, also included "t.me/", the prefix for any invitation to Telegram groups and channels, "https://x.com/search?q=lang%3Apt%20%22t.me%2F%22%20SEARCH-TERM".

With the implementation of community identification approaches for conspiracy theories developed over months of investigation and method refinement, it was possible to build a project database encompassing a total of 855 Brazilian conspiracy theory communities on Telegram (including other themes not covered in this study). These communities have collectively published 27,227,525 pieces of content from May 2016 (the first publications) to August 2024 (the period of this study), with a combined total of 2,290,621 users across the Brazilian communities. It is important to note that this volume of users includes two elements: first, it is a variable figure, as users can join and leave communities daily, so this value represents what was recorded on the publication extraction date; second, it is possible that the same user is a member of more than one group and, therefore, is counted more than once. In this context, while the volume remains significant, it may be slightly lower when considering the deduplicated number of citizens within these Brazilian conspiracy theory communities.

## 2.2. Data processing

With all the Brazilian conspiracy theory groups and channels on Telegram extracted, a manual classification was conducted considering the title and description of the community. If there was an explicit mention in the title or description of the community related to a specific theme, it was classified into one of the following categories: (i) "Anti-Science"; (ii) "Anti-Woke and Gender"; (iii) "Antivax"; (iv) "Apocalypse and Survivalism"; (v) "Climate Changes"; (vi) "Flat Earth"; (vii) "Globalism"; (viii) "New World Order"; (ix) "Occultism and Esotericism"; (x) "Off Label and Quackery"; (xi) "QAnon"; (xii) "Reptilians and Creatures"; (xiii) "Revisionism and Hate Speech"; (xiv) "UFO and Universe". If there was no explicit mention related to the themes in the title or description of the community, it was classified as (xv) "General Conspiracy". In the following table, we can observe the metrics related to the classification of these conspiracy theory communities in Brazil.

**Table 01.** Conspiracy theory communities in Brazil (metrics up to August 2024)

| Categories | Groups | Users | Contents | Comments | Total |
|---|---|---|---|---|---|
| **Anti-Science** | 22 | 58,138 | 187,585 | 784,331 | 971,916 |
| **Anti-Woke and Gender** | 43 | 154,391 | 276,018 | 1,017,412 | 1,293,430 |
| **Antivax** | 111 | 239,309 | 1,778,587 | 1,965,381 | 3,743,968 |
| **Apocalypse and Survivalism** | 33 | 109,266 | 915,584 | 429,476 | 1,345,060 |
| **Climate Changes** | 14 | 20,114 | 269,203 | 46,819 | 316,022 |
| **Flat Earth** | 33 | 38,563 | 354,200 | 1,025,039 | 1,379,239 |
| **General Conspiracy** | 127 | 498,190 | 2,671,440 | 3,498,492 | 6,169,932 |
| **Globalism** | 41 | 326,596 | 768,176 | 537,087 | 1,305,263 |
| **NWO** | 148 | 329,304 | 2,411,003 | 1,077,683 | 3,488,686 |
| **Occultism and Esotericism** | 39 | 82,872 | 927,708 | 2,098,357 | 3,026,065 |
| **Off Label and Quackery** | 84 | 201,342 | 929,156 | 733,638 | 1,662,794 |
| **QAnon** | 28 | 62,346 | 531,678 | 219,742 | 751,420 |
| **Reptilians and Creatures** | 19 | 82,290 | 96,262 | 62,342 | 158,604 |
| **Revisionism and Hate Speech** | 66 | 34,380 | 204,453 | 142,266 | 346,719 |
| **UFO and Universe** | 47 | 58,912 | 862,358 | 406,049 | 1,268,407 |
| **Total** | **855** | **2,296,013** | **13,183,411** | **14,044,114** | **27,227,525** |

Source: Own elaboration (2024).

With this volume of extracted data, it was possible to segment and present in this study only communities and content related to flat earth themes. In parallel, other themes of Brazilian conspiracy theory communities were also addressed with studies aimed at characterizing the extent and dynamics of the network, which are openly and originally available on arXiv at Cornell University.

Additionally, it should be noted that only open communities were extracted, meaning those that are not only publicly identifiable but also do not require any request to access the content, being available to any user with a Telegram account who needs to join the group or channel. Furthermore, in compliance with Brazilian legislation, particularly the General Data Protection Law (LGPD), or Law No. 13,709/2018 (Brazilian law from 2018), which deals with privacy control and the use/treatment of personal data, all extracted data were anonymized for the purposes of analysis and investigation. Therefore, not even the identification of the communities is possible through this study, thus extending the user's privacy by anonymizing their data beyond the community itself to which they submitted by being in a public and open group or channel on Telegram.

### 2.3. Approaches to data analysis

A total of 30 selected communities focused on flat earth themes, containing 632,901 publications and 38,520 combined users, will be analyzed. Four approaches will be used to investigate the conspiracy theory communities selected for the scope of this study. These metrics are detailed in the following table:

**Table 02.** Selected communities for analysis (metrics up to August 2024)

| Category | Groups | Users | Contents | Comments | Total |
|---|---|---|---|---|---|
| **Flat earth** | 30 | 38,520 | 188,068 | 444,833 | 632,901 |

Source: Own elaboration (2024).

**(i) Network:** by developing a proprietary algorithm to identify messages containing the term "t.me/" (inviting users to join other communities), we propose to present volumes and connections observed on how **(a)** communities within the flat earth theme circulate invitations for their users to explore more groups and channels within the same theme, reinforcing shared belief systems; and how **(b)** these same communities circulate invitations for their users to explore groups and channels dealing with other conspiracy theories, distinct from their explicit purpose. This approach is valuable for observing whether these communities use themselves as a source of legitimation and reference and/or rely on other conspiracy theory themes, even opening doors for their users to explore other conspiracies. Furthermore, it is worth mentioning the study by Rocha *et al.* (2024), where a network identification approach was also applied in Telegram communities, but by observing similar content based on an ID generated for each unique message and its similar ones;

**(ii) Time series:** the "Pandas" library (McKinney, 2010) is used to organize the investigation data frames, observing **(a)** the volume of publications over the months; and **(b)** the volume of engagement over the months, considering metadata of views, reactions, and comments collected during extraction. In addition to volumetry, the "Plotly" library (Plotly Technologies Inc., 2015) enabled the graphical representation of this variation;

**(iii) Content analysis:** in addition to the general word frequency analysis, time series are applied to the variation of the most frequent words over the semesters—observing from June 2019 (initial publications) to August 2024 (when this study was conducted). With the "Pandas" (McKinney, 2010) and "WordCloud" (Mueller, 2020) libraries, the results are presented both volumetrically and graphically;

**(iv) Thematic agenda overlap:** following the approach proposed by Silva & Sátiro (2024) for identifying thematic agenda overlap in Telegram communities, we used the "BERTopic" model (Grootendorst, 2020). BERTopic is a topic modeling algorithm that facilitates the generation of thematic representations from large amounts of text. First, the algorithm extracts document embeddings using sentence transformer models, such as "all-MiniLM-L6-v2". These embeddings are then reduced in dimensionality using techniques like "UMAP", facilitating the clustering process. Clustering is performed using "HDBSCAN", a density-based technique that identifies clusters of different shapes and sizes, as well as outliers. Subsequently, the documents are tokenized and represented in a bag-of-words structure, which is normalized (L1) to account for size differences between clusters. The topic representation is refined using a modified version of "TF-IDF", called "Class-TF-IDF", which considers the importance of words within each cluster (Grootendorst, 2020). It is important to note that before applying BERTopic, we cleaned the dataset by removing Portuguese "stopwords" using "NLTK" (Loper & Bird, 2002). For topic modeling, we used the "loky" backend to optimize performance during data fitting and transformation.

In summary, the methodology applied ranged from data extraction using the own tool TelegramScrap (Silva, 2023) to the processing and analysis of the collected data, employing various approaches to identify and classify Brazilian conspiracy theory communities on Telegram. Each stage was carefully implemented to ensure data integrity and respect for user privacy, as mandated by Brazilian legislation. The results of this data will be presented below, aiming to reveal the dynamics and characteristics of the studied communities.

## 3. Results

The results are detailed below in the order outlined in the methodology, beginning with the characterization of the network and its sources of legitimation and reference, progressing to the time series, incorporating content analysis, and concluding with the identification of thematic agenda overlap among the conspiracy theory communities.

### 3.1. Network

The analysis of the network of flat-earth communities provides us with a detailed view of how these theories are disseminated and interconnected within the conspiratorial ecosystem. The figures presented explore different aspects of these connections, revealing both the internal structure of flat-earth networks and their role as gateways to and from other conspiracy theories. When analyzing the internal network (Figure 01), we can observe that the dissemination of flat-earth beliefs is largely centralized, with a few communities acting as central hubs where interactions and content sharing are concentrated. This centralization suggests that a small number of groups exert significant influence in perpetuating these narratives, while smaller communities, although peripheral, are still impacted by the centralized content.

In Figures 02 and 03, which address the gateway networks, it is evident that flat-earth beliefs are connected to other conspiracy theories, such as the New World Order and Occultism. This indicates that flat-earth beliefs can both introduce new adherents to a broader conspiratorial universe and serve as a starting point for individuals to migrate to more complex and diverse discussions. This interconnectedness is essential for understanding how beliefs surrounding the flat Earth can facilitate the transition to other theories, expanding individuals' engagement within these communities.

Finally, the flow of invitation links between flat-earth communities (Figure 04) reveals the dynamics of recruitment and radicalization that occur within these networks. Flat-earth beliefs, when linked to themes such as the New World Order and Apocalypse, function as a gateway to deeper and more ideologically charged narratives. This figure illustrates how flat-earth adherents are directed towards other central conspiracy theories, reinforcing the role of flat-earth beliefs as an ideological commitment test that, once passed, leads individuals to deeper involvement with large-scale theories. Thus, flat-earth beliefs not only maintain their follower base but also play a crucial role in integrating new members into a broader and more interconnected network of conspiratorial beliefs.

**Figure 01.** Internal network between flat earth communities

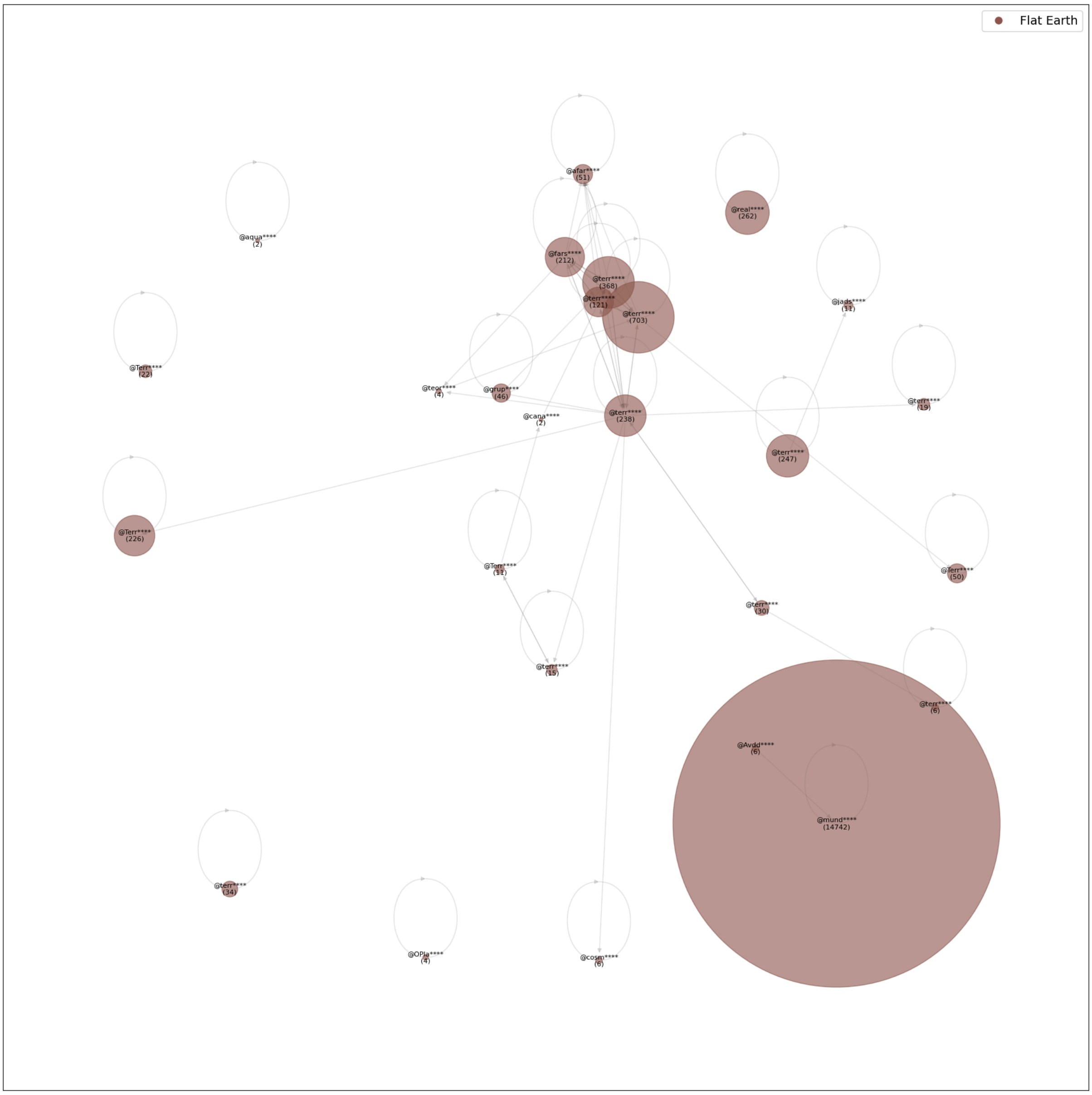

Source: Own elaboration (2024).

The figure illustrates an internal network among communities that promote flat-earth theories. The connections reveal a structure where a few communities act as central hubs, with the largest nodes indicative of the main sources of dissemination for this theory. The network, though less dense compared to others, suggests a centralization in the propagation of information, with a strong interconnection between larger groups and smaller ones that orbit around them. These central hubs appear to concentrate most of the interactions and content sharing, indicating that a limited number of groups exert significant influence in perpetuating flat-earth narratives. The presence of more isolated peripheral communities also suggests that, although some sub-communities may not have a direct connection to the central core, they are still impacted by the content disseminated by the larger influencers within the network.

**Figure 02.** Network of communities that open doors to the theme (gateway)

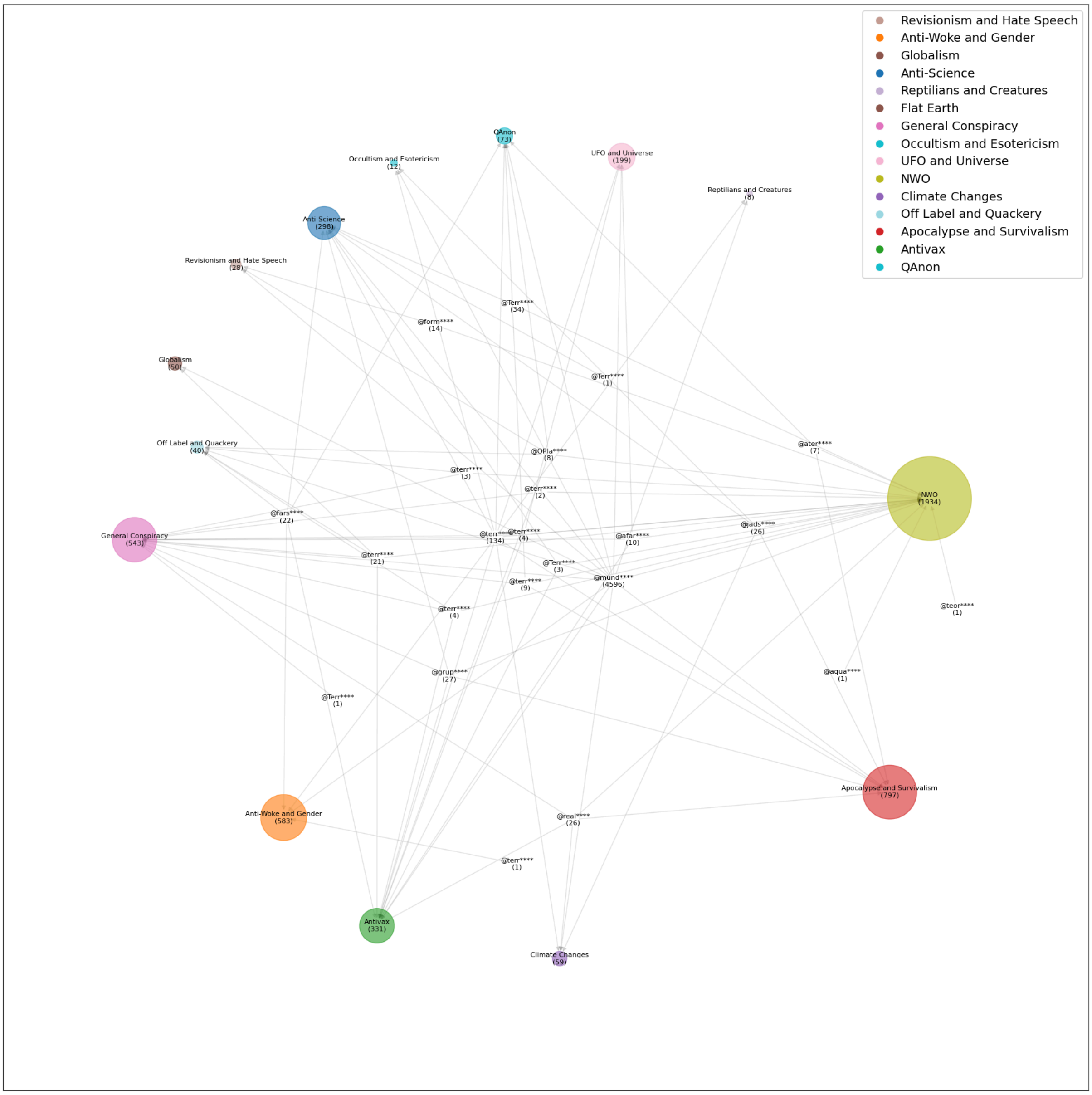

Source: Own elaboration (2024).

The figure illustrates the network of communities that serve as gateways to flat-earth beliefs. It is observed that, although flat-earth beliefs are a relatively isolated conspiracy theory, they are still connected to other communities discussing similar themes, such as the New World Order and Occultism. These connections indicate that flat-earth beliefs can be a gateway to broader theories or vice versa. The graph suggests that once individuals are exposed to flat-earth beliefs, there is a likelihood of migration to other conspiracy theories, with larger communities acting as hubs that centralize and disseminate related content, facilitating the transition to other conspiratorial areas.

**Figure 03.** Network of communities whose theme opens doors (exit point)

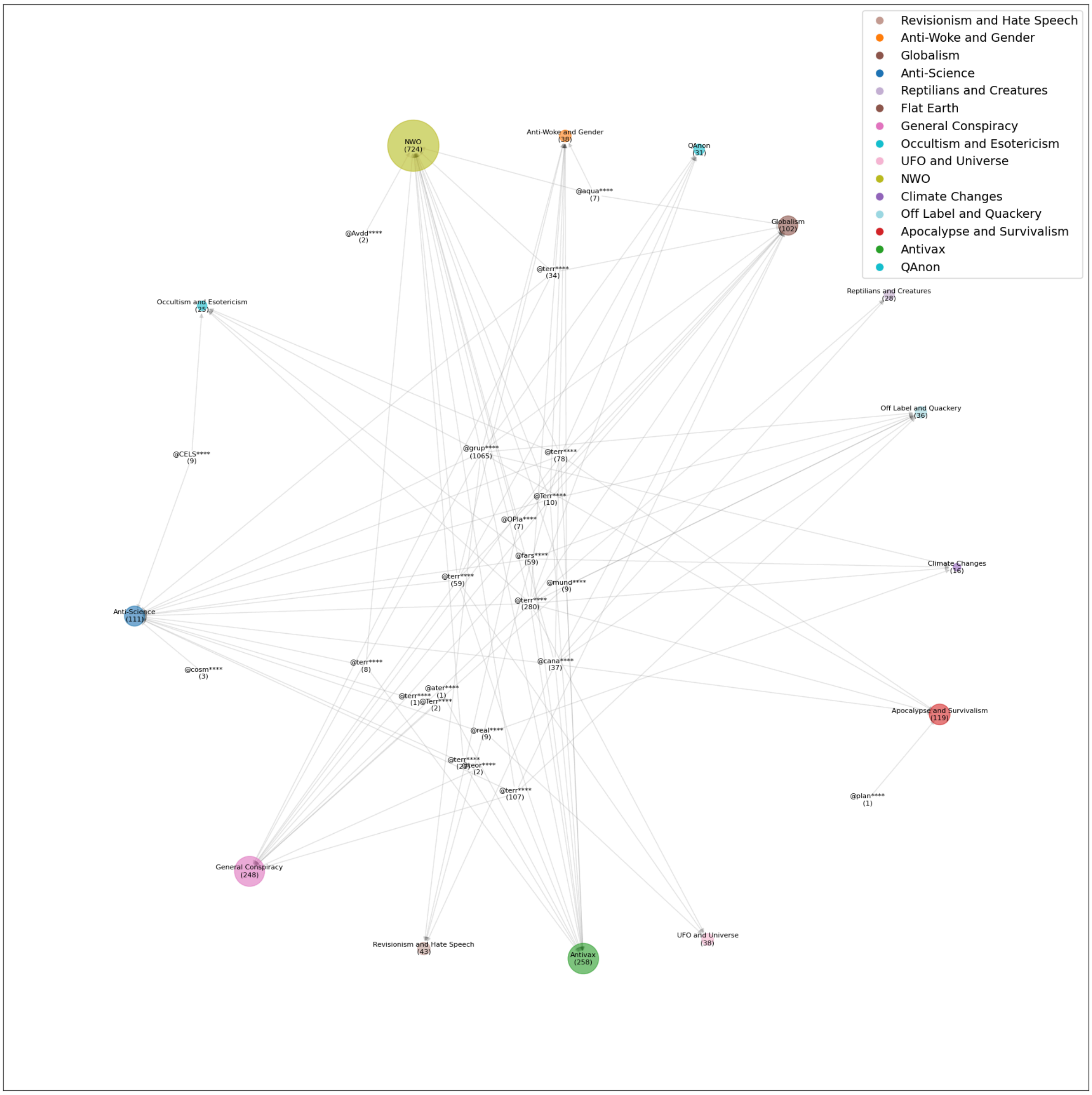

Source: Own elaboration (2024).

This graph reveals the interconnections between communities centered on the flat-earth theory and how these communities can act as catalysts for the exploration of other conspiracy theories. It is observed that, although the flat-earth theory is a specific niche within the conspiratorial universe, the communities associated with it have connections with a diverse range of other themes. The presence of large nodes, such as “New World Order” and “Occultism and Esotericism”, suggests that followers of the flat-earth theory can easily transition to broader and more complex discussions, expanding their involvement within the conspiratorial ecosystem. The figure illustrates how the flat-earth theory can serve as an "exit gateway" to other theories, showing that once individuals are embedded in this niche, they are frequently exposed to a broader universe of related theories, often with a distorted and amplified worldview.

**Figure 04.** Flow of invitation links between flat earth communities

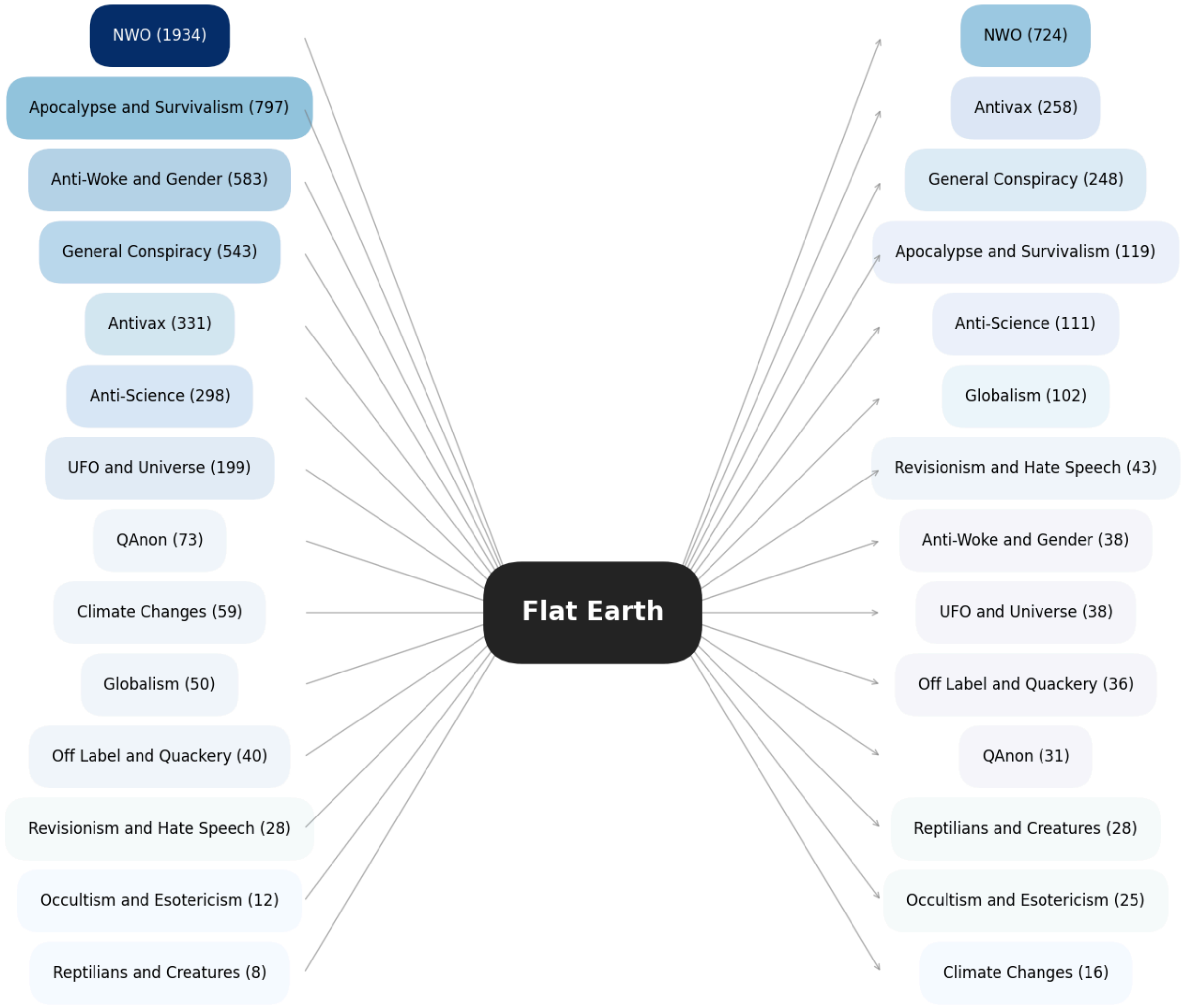

Source: Own elaboration (2024).

The flowchart of invitation links related to flat-earth beliefs reveals a scenario where this theory, often treated as a marginal and exotic topic, acts as a significant gateway to a broader ecosystem of conspiracy theories. The predominance of links from the New World Order (1,934) and themes like Apocalypse and Survivalism (797) and Anti-Woke and Gender (583) suggests that flat-earth beliefs serve as an initial access layer to more complex and profound narratives. This dynamic is crucial for understanding the role that flat-earth beliefs play in recruiting and radicalizing individuals within these networks. The flat-earth movement is not just an isolated belief; it is an anchoring point that prepares adherents to embrace broader narratives of resistance against established scientific knowledge and the global order.

In the opposite direction, with NWO receiving 724 links from flat-earth beliefs, we observe that adherents of this theory are quickly directed towards more centralized narratives of global control and resistance to science. This reflects how flat-earth beliefs can be instrumentalized to engage individuals in broader and ideologically charged discourses, indicating that this theory functions as a sort of ideological commitment test that, once passed, leads individuals to deeper involvement with large-scale theories.

### 3.2. Time series

The analysis of time series allows us to observe how flat-earth beliefs, a conspiracy theory that challenges established scientific foundations, gained traction at key moments in recent years. In the following graph, we see how mentions of this theory abruptly increased in 2020, possibly fueled by the uncertainties generated by the COVID-19 pandemic and subsequent global restrictions. This surge was followed by a second peak in January 2021, coinciding with a period of heightened political distrust in the United States. These peaks reveal how global and political crises can catalyze the spread of conspiracy theories, creating a fertile environment for the proliferation of radical ideas. As we move forward in time, a stabilization is noted, indicating a possible normalization of these topics within digital culture.

**Figure 05.** Line graph over the period

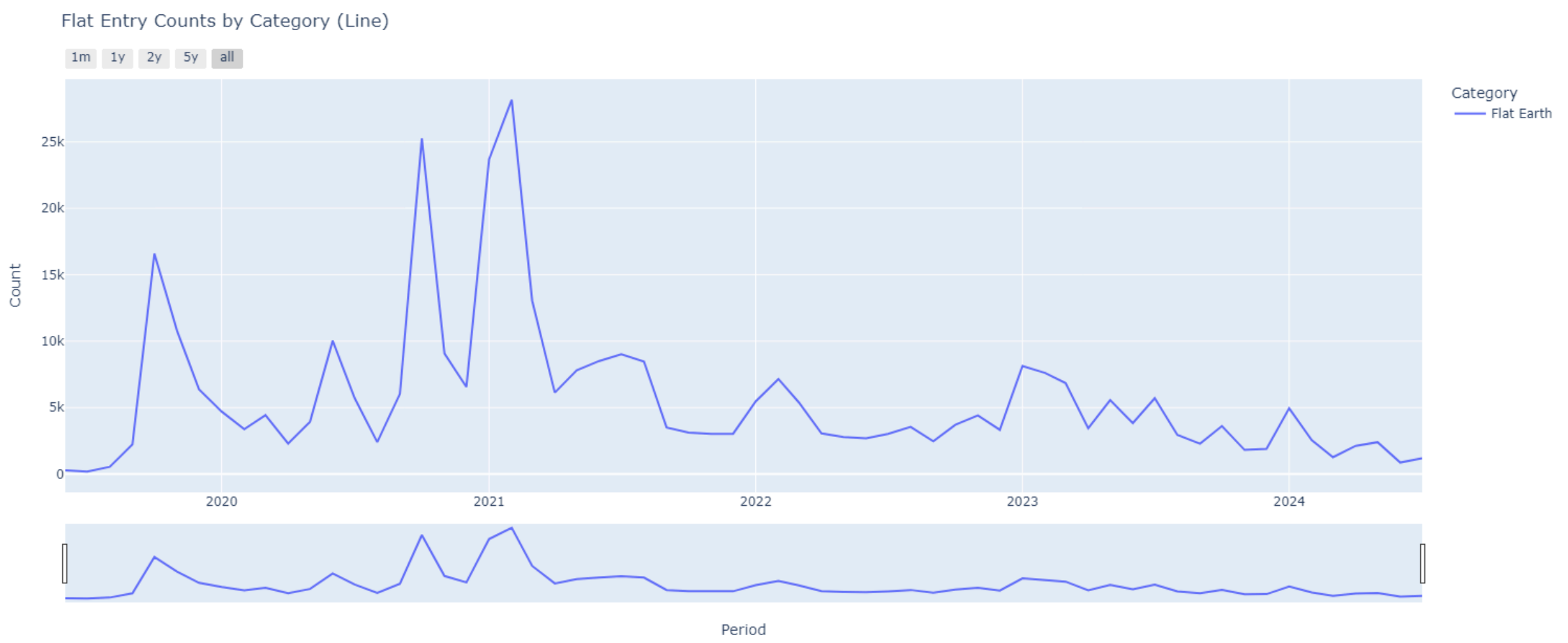

Source: Own elaboration (2024).

In the flat-earth graph, we observe that the first significant peak occurs in mid-2020, where mentions increase from around 500 to approximately 2,500, representing a 400% growth in a short period. This surge coincides with the onset of global mobility restrictions due to COVID-19, when conspiracy theories about "manipulation" by science began to proliferate. Another relevant peak occurs in January 2021, where mentions again reach around 2,500. Compared to the initial volume in 2020, this represents a variation of about 300%. This peak can be associated with the climate of distrust generated by the U.S. presidential elections, where various types of conspiracy theories, including flat-earth beliefs, intertwined with political discussions. After these peaks, starting in 2022, we see a stabilization around 1,000 to 1,500 mentions per month, a decrease of approximately 40% compared to previous peaks. This suggests a reduction in the intensity of discussions, despite ongoing engagement.

### 3.3. Content analysis

The content analysis of flat-earth communities through word clouds provides a detailed view of how certain terms and concepts consolidate and evolve within these discussions over time. From the most frequent words, such as “earth”, “flat”, “truth”, and “group”. the central foundations of flat-earth discourse emerge, revealing a linguistic pattern that reflects both the attempt to reaffirm conspiratorial beliefs and the resistance to conventional science. This analysis allows us to observe how flat-earth beliefs not only challenge the shape of the Earth but are also closely linked to a broader worldview that includes religious elements, as indicated by the presence of the term “God”, and elements of distrust towards institutions and authorities. The emphasis on words like “channel” and “Telegram” over the years suggests a focus on disseminating these ideas through digital platforms, demonstrating the importance of these media in the perpetuation and expansion of flat-earth beliefs. Thus, the word cloud not only highlights the main themes discussed but also reveals how these communities organize and communicate to sustain and spread convictions.

**Figure 06.** Consolidated word cloud for flat earth medications

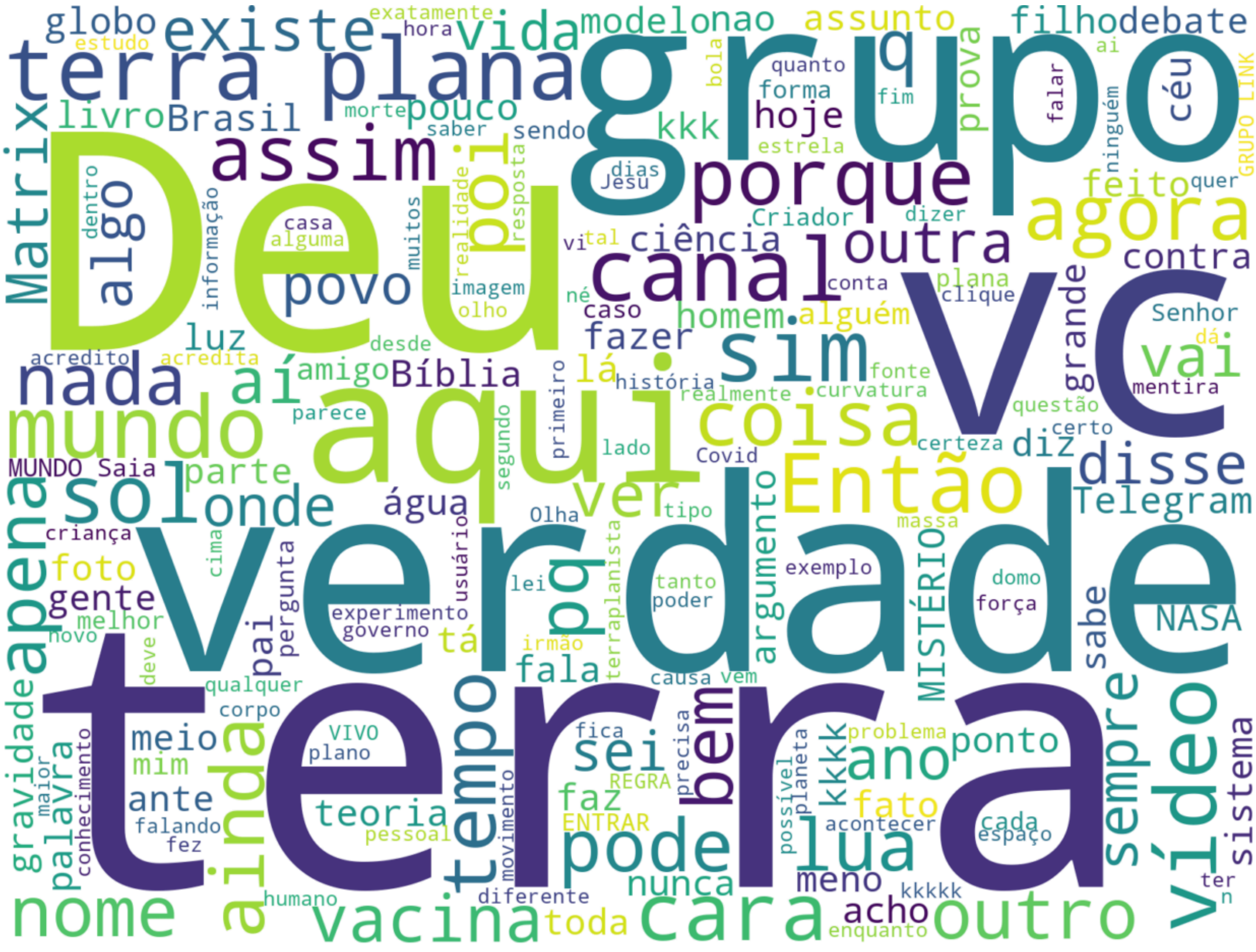

Source: Own elaboration (2024).

[ english version / português abaixo ]

The consolidated word cloud of flat-earth beliefs reveals the centrality of terms such as “earth”, “flat”, “truth”, “group”, and “now”. The prominence given to the word “truth” suggests a continuous effort by members of these communities to assert an alternative reality, where claims about the shape of the Earth are taken as an absolute truth that opposes conventional scientific knowledge. The word “group” underscores the crucial role of communities in disseminating these beliefs, functioning as spaces for mutual reaffirmation and resistance against the dominant perception. Additionally, the presence of terms like “God”, “life”, and “world” indicates that discussions within these communities are not limited to scientific issues but extend to spiritual and existential aspects, suggesting that flat-earth beliefs are part of a broader worldview that combines religious and conspiratorial elements. The recurrence of the word “channel” points to the importance of digital media, such as YouTube and Telegram, in propagating these ideas, where specific videos and groups play a central role in maintaining and expanding the community.

**Chart 01.** Temporal word cloud series for flat earth narratives

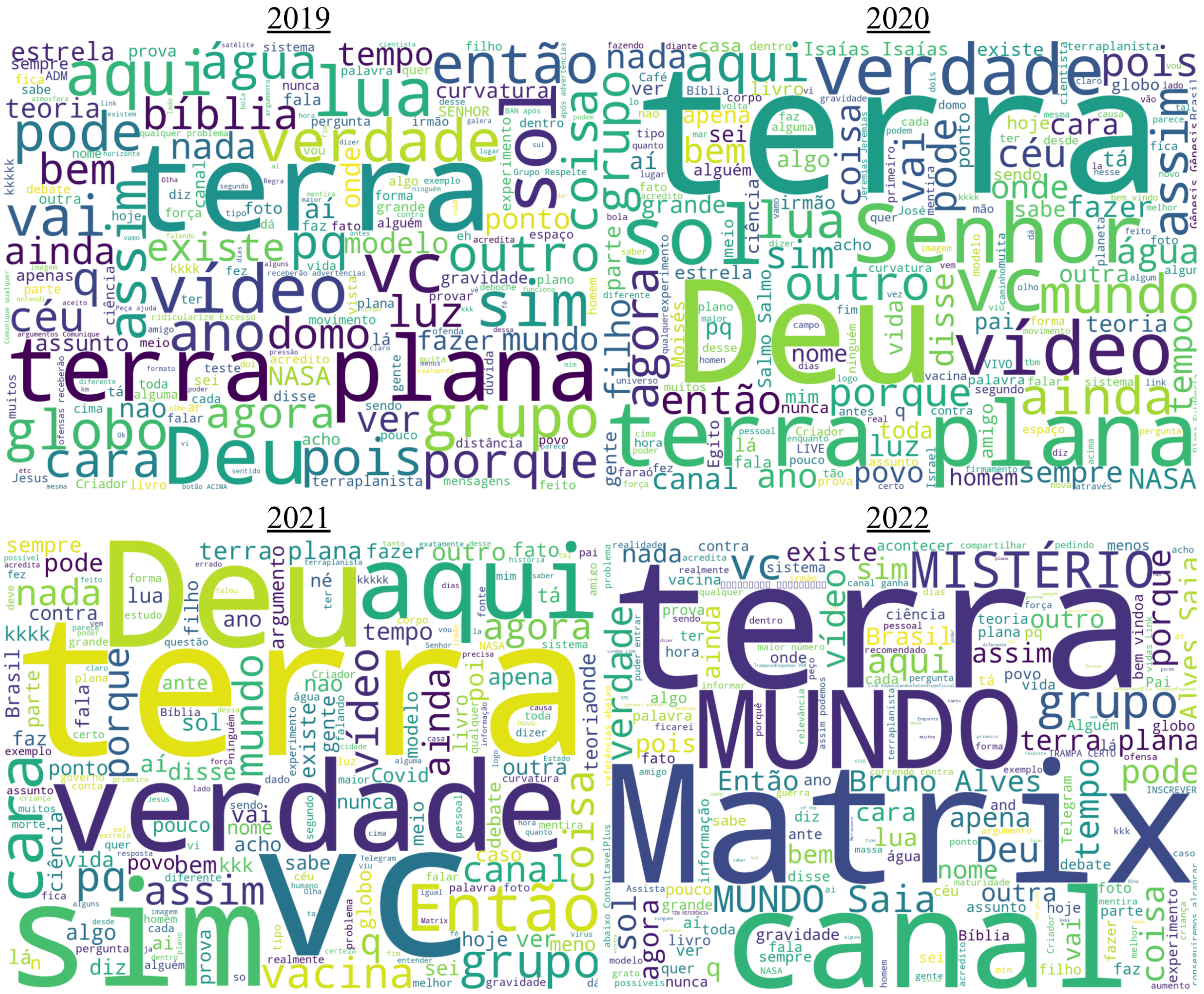

[ english version / português abaixo ]

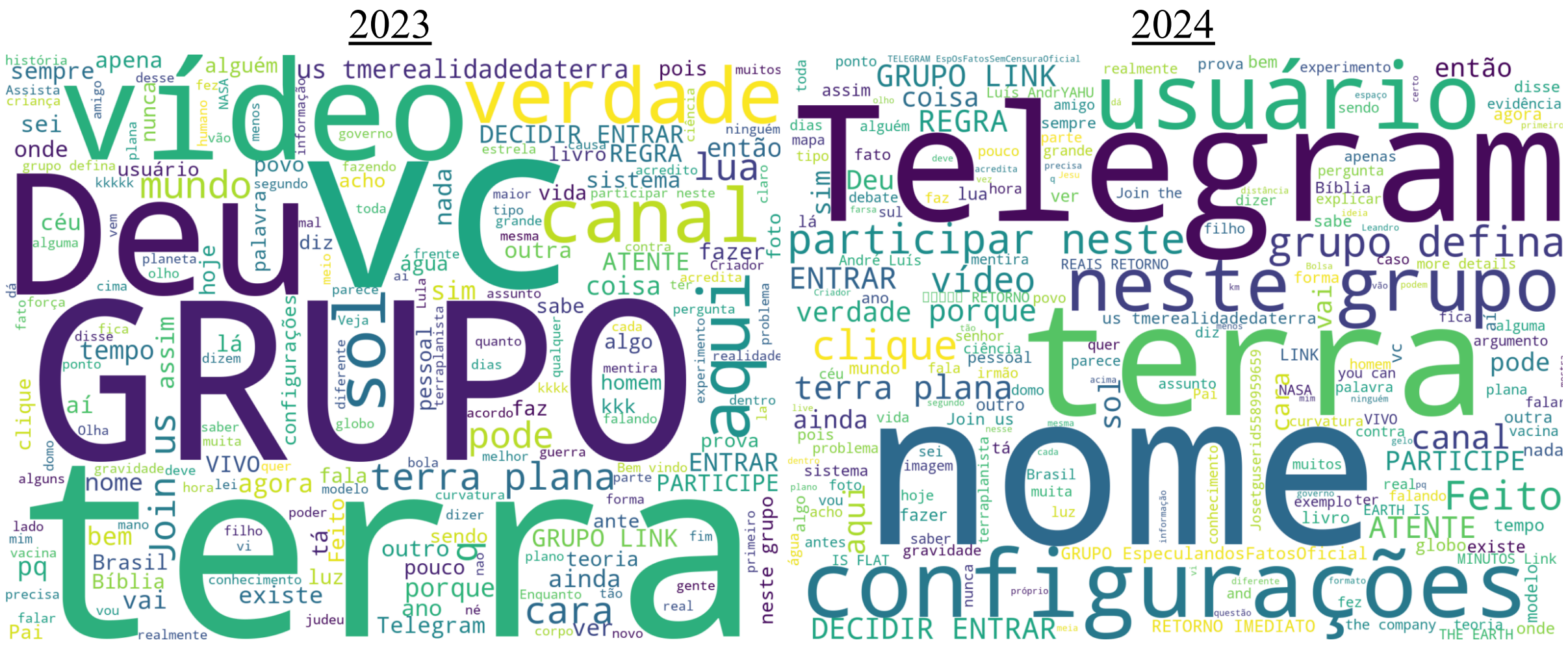

Source: Own elaboration (2024).

The time series word cloud for flat-earth beliefs reveals how the discussions and priorities of these communities have evolved over the years. In 2019, the words “earth” and “flat” already dominated the discussions, indicating a phase of affirmation of these basic ideas within the community. The presence of terms like “moon” and “energy” also suggests an interest in theories that connect flat-earth beliefs to more esoteric concepts. In 2020, with the impact of the COVID-19 pandemic, the entry of terms like “Matrix” and “world” is observed, which may reflect an increase in discussions about control and simulated reality, often linked to the idea that the pandemic is part of a larger plan of global manipulation. In 2021, “truth” and “channel” gain prominence, suggesting an intensification in the dissemination of content that seeks to discredit official science and promote alternative narratives through digital media channels. In the following years, especially in 2023 and 2024, there is a greater diversification of themes, with the inclusion of terms like “Telegram” and “settings”, which may indicate an adaptation of communication strategies, perhaps in response to censorship or restrictions on other platforms. This evolution highlights how flat-earth communities not only persist but also adapt to the social and technological context, ensuring narrative expansion.

### 3.4. Thematic agenda overlap

The following figures represent an analysis of thematic agenda overlap in conspiracy theory communities, with a particular focus on flat-earth beliefs. By visualizing the topics discussed by these communities, it is possible to identify how different disinformation narratives are interconnected, creating a cohesive discourse that reinforces the core beliefs of these communities. This analysis aims to highlight the connections between conspiracy theories and how seemingly distinct themes, such as science, religion, and geopolitics, are used to sustain the flat-earth narrative.

**Figure 07.** Flat-earth and rejection of gravity based on the Bible themes

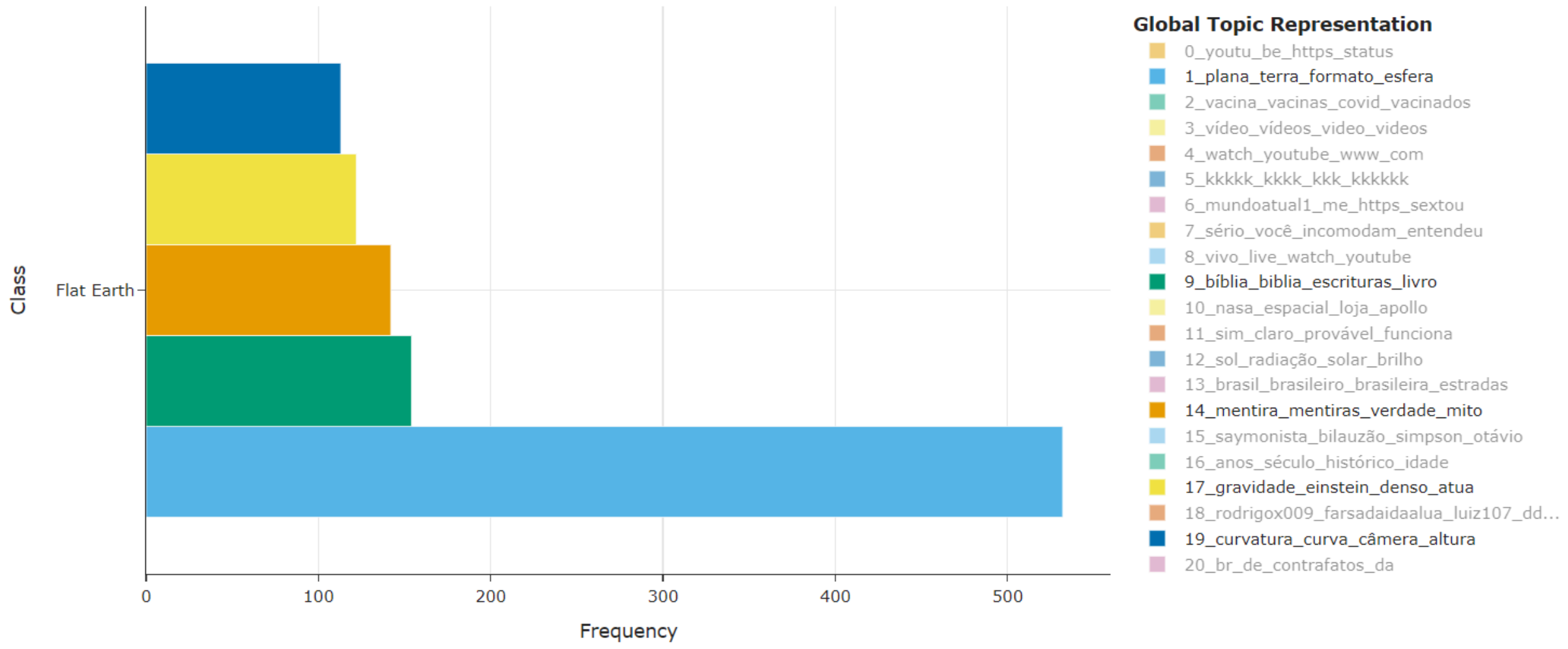

Source: Own elaboration (2024).

Figure 07 explores the relationship between flat-earth beliefs and the rejection of gravity based on biblical interpretations. The predominance of topics such as “Bible”, “scriptures”, and “book” indicates that these communities use religious passages to justify their beliefs that gravity is an invented concept designed to deceive humanity. The overlap of topics related to religion with the flat-earth theory suggests an attempt to legitimize this belief through a literal reading of the Bible, creating a connection between science and faith that resists rational arguments based on evidence.

**Figure 08.** Faith and demons linked to the flat-earth narrative themes

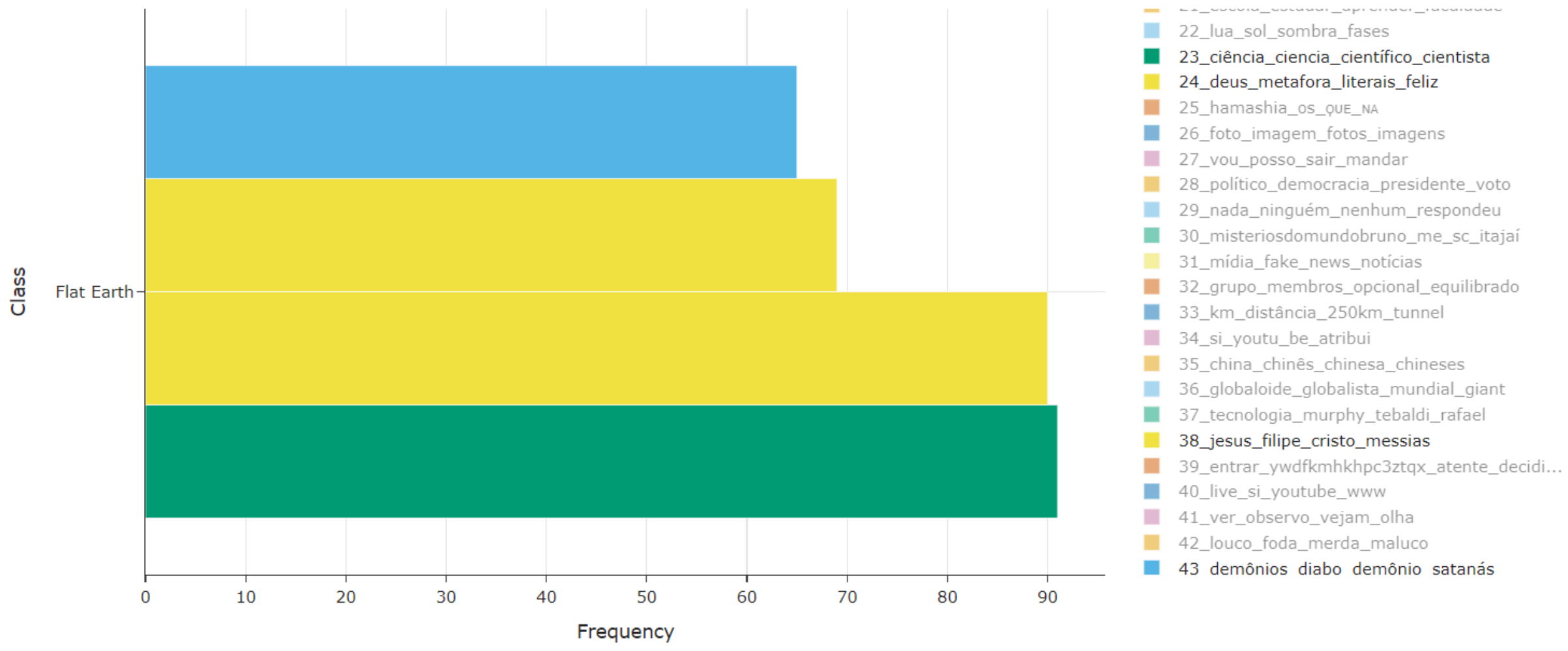

Source: Own elaboration (2024).

In Figure 08, the connection between flat-earth beliefs and themes of faith and demons is evident. Topics such as “demons”, “devil”, and “Satan” indicate that discussions in these communities often associate belief in a flat Earth with a spiritual battle, where the acceptance of modern science is viewed as a demonic act. This approach strengthens the idea that belief in a flat Earth is part of a spiritual resistance against evil forces, consolidating flat-earth beliefs as something that transcends scientific debate to become a moral and religious issue.

**Figure 09.** Criticism of species evolution based on the Bible themes

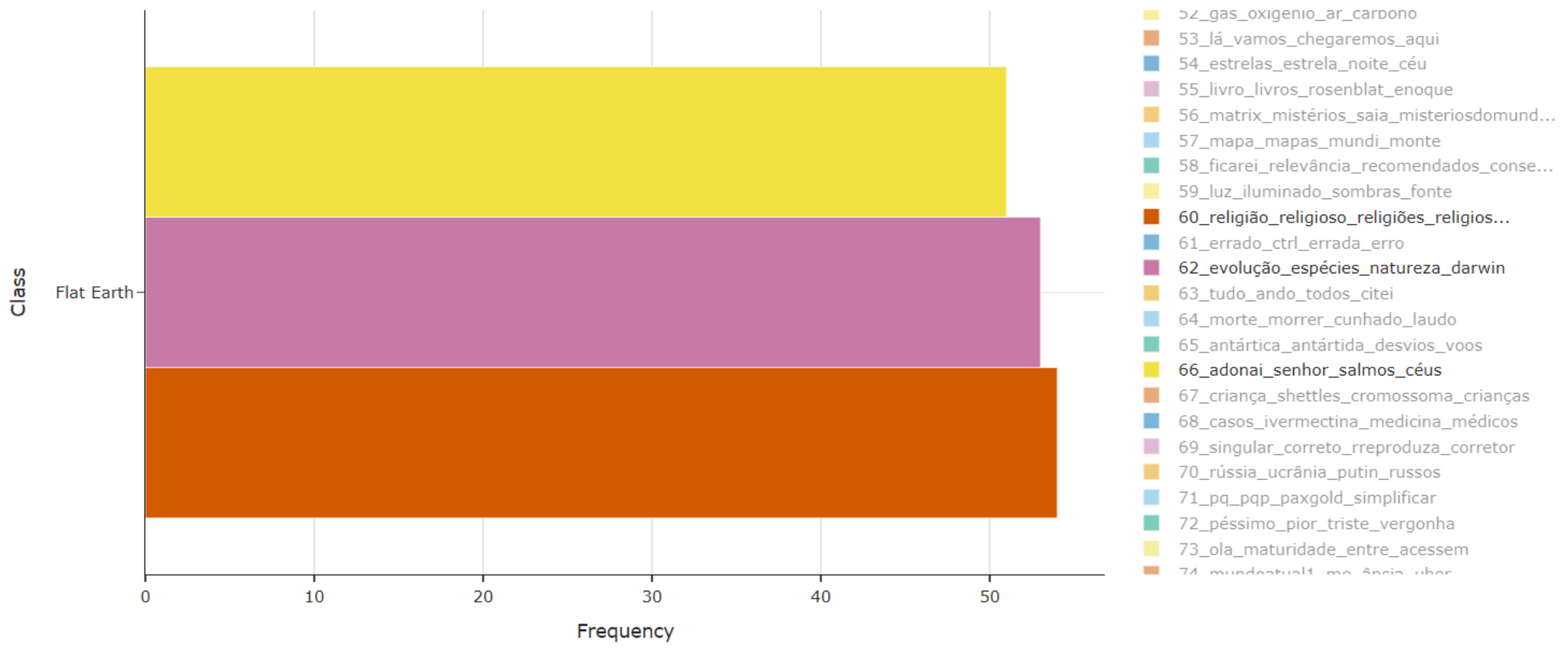

Source: Own elaboration (2024).

Figure 09 highlights the criticism of the theory of species evolution, supported by biblical arguments within flat-earth communities. Topics such as “evolution”, “species”, and “Darwin” are discussed in ways that deny the scientific validity of evolution, using religious interpretations to support this rejection. This overlap of themes reflects an effort to discredit widely accepted scientific theories, reinforcing a worldview that positions the biblical narrative as the only unquestionable truth, directly opposing scientific evidence.

**Figure 10.** Solar radiation and solar flares themes

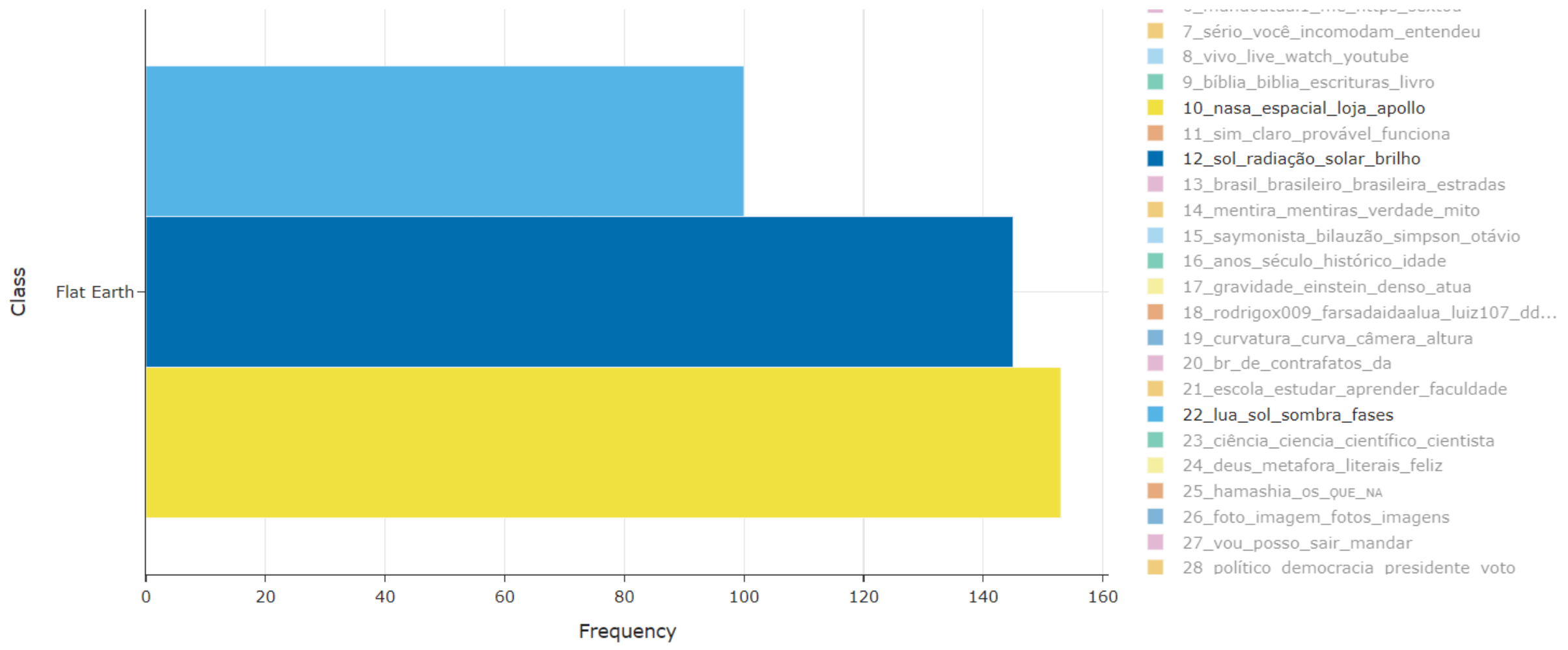

Source: Own elaboration (2024).

In Figure 10, the theme of solar radiation and solar flares is explored in connection with flat-earth beliefs. Topics such as “sun”, “radiation”, and “solar flares” are discussed in ways that suggest astronomical events are manipulated or misinterpreted by traditional science. This approach serves to reinforce distrust towards space and cosmological science, supporting the idea that flat Earth is the true representation of reality, while phenomena like solar radiation are seen as part of a hoax.

**Figure 11.** Biblical apocalypse and pseudofrequencies themes

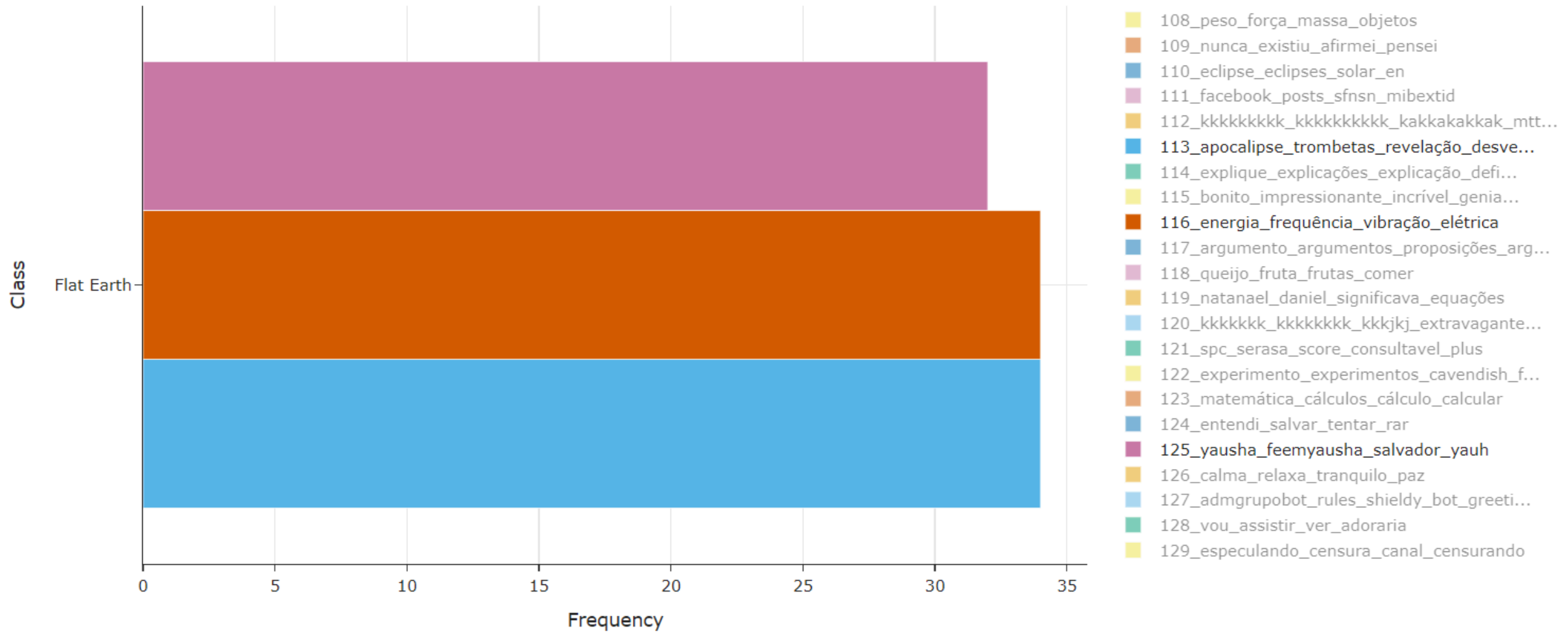

Source: Own elaboration (2024).

Figure 11 explores the intersection between flat-earth beliefs and biblical apocalypse narratives, along with discussions on pseudofrequencies. Topics such as “apocalypse”, “trumpets”, and “vibration” show that these communities often connect their flat-earth beliefs with apocalyptic predictions based on the Bible. This intersection of themes suggests that flat-earth beliefs are seen not only as a belief about the shape of the Earth but as part of a broader worldview of an imminent end, where cosmic events are interpreted as divine signs.

**Figure 12.** Magnetic pole reversal and cartographic denial themes

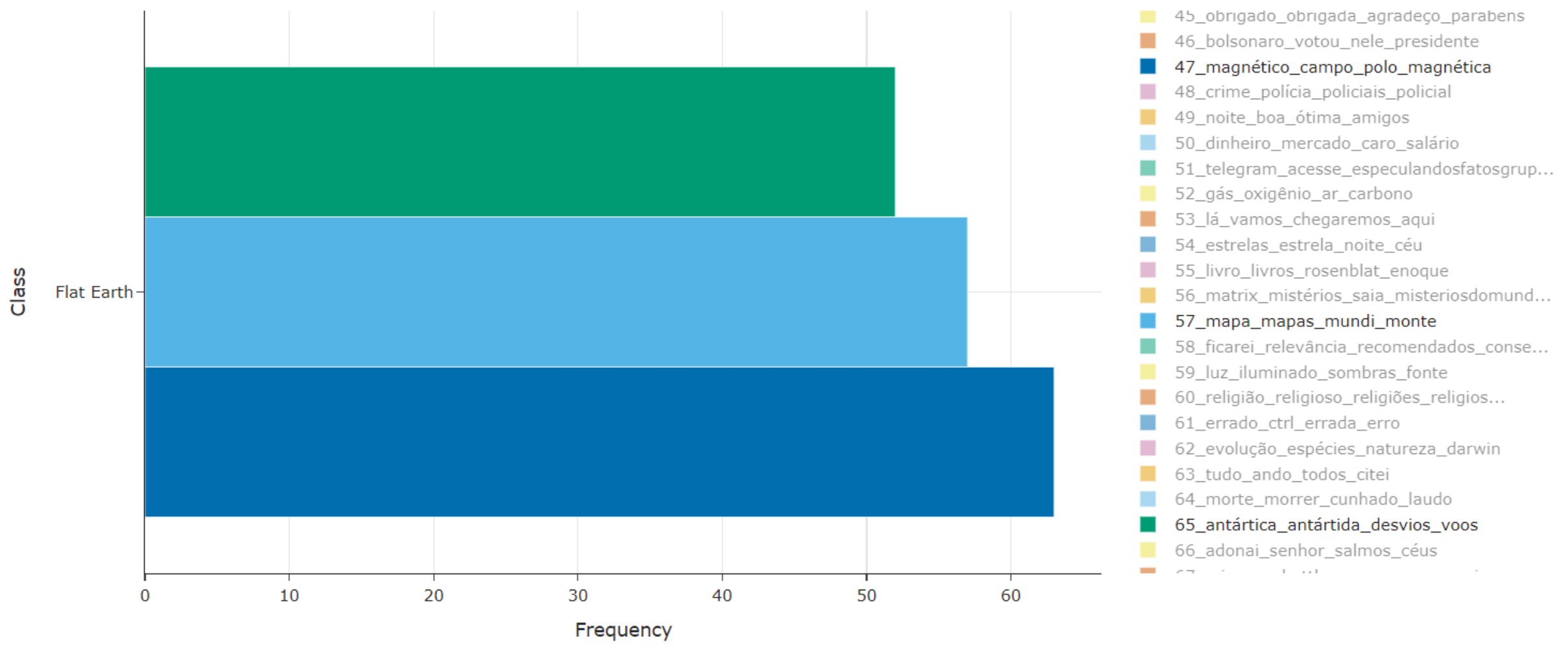

Source: Own elaboration (2024).

In Figure 12, themes related to magnetic pole reversal and cartographic denial are discussed within the context of flat-earth beliefs. Topics such as “magnetic pole”, “maps”, and "cartography" are addressed to question conventional scientific interpretations of Earth's geography and magnetic phenomena. This denial of cartographic and geophysical principles reflects the attempt of these communities to construct a new narrative about the physical reality of the Earth, discrediting the foundations of modern geography and supporting the belief in a flat Earth.

## 4. Reflections and future works

To answer the research question, **"how are Brazilian conspiracy theory communities on flat earth topics characterized and articulated on Telegram?"**, this study adopted mirrored techniques in a series of seven publications aimed at characterizing and describing the phenomenon of conspiracy theories on Telegram, focusing on Brazil as a case study. After months of investigation, it was possible to extract a total of 30 Brazilian conspiracy theory communities on Telegram focused on flat earth topics, amounting to 632,901 pieces of content published between June 2019 (initial publications) and August 2024 (when this study was conducted), with 38,520 users combined across the communities.

Four main approaches were adopted: **(i)** Network, which involved the creation of an algorithm to map connections between communities through invitations circulated among groups and channels; **(ii)** Time series, which used libraries like "Pandas" (McKinney, 2010) and "Plotly" (Plotly Technologies Inc., 2015) to analyze the evolution of publications and engagements over time; **(iii)** Content analysis, where textual analysis techniques were applied to identify patterns and word frequencies in the communities over the semesters; and **(iv)** Thematic agenda overlap, which utilized the BERTopic model (Grootendorst, 2020) to group and interpret large volumes of text, generating coherent topics from the analyzed publications. The main reflections are detailed below, followed by suggestions for future works.

### 4.1. Main reflections

**Significant increase in flat-earth discussions during the COVID-19 pandemic:** During the peak of the pandemic, discussions in flat-earth communities on Brazilian Telegram significantly increased, with a 400% growth in the volume of mentions of this theory between 2019 and 2020. This surge coincided with the spread of disinformation related to the pandemic, suggesting that flat-earth beliefs were largely driven by a widespread environment of distrust in scientific and governmental institutions;

**Flat-earth as a gateway to other conspiracy theories:** Flat-earth communities not only attract individuals to their ranks but also act as portals for their members to be exposed to other conspiracy theories, such as the New World Order and Occultism. With a total of 1,934 links from NWO and 797 from Apocalypse and Survivalism, it is clear that these theories are strongly interconnected, facilitating an ideological transition between different disinformative narratives;

**Centralization of discussions in a few communities:** The network of flat-earth communities is highly centralized, with a small number of groups acting as the main hubs for content dissemination. These central communities possess a disproportionate capacity to influence discourse, serving as epicenters for the perpetuation and amplification of flat-earth narratives, while smaller groups orbit around them;

**Thematic interconnectivity with the New World Order:** Discussions about the New World Order are among the most frequently mentioned in flat-earth communities, with 724 links identified as directed towards this theme. This suggests that the narrative of a supposed global control is often used to legitimize and reinforce flat-earth beliefs, creating thematic overlap that strengthens these narratives;

**Role of digital media in the spread of flat-earth beliefs:** Terms such as “channel” and “Telegram” were identified as central in discussions, indicating that the spread of flat-earth beliefs relies heavily on digital platforms, guiding mobilizations. The prevalence of these terms over the years highlights the importance of Telegram and other platforms in the maintenance and expansion of flat-earth communities;

**Relationship between flat-earth beliefs and religious narratives:** Content analysis revealed a strong presence of religious terms, such as “God“ and “Bible”, in discussions about flat-earth beliefs. This indicates that flat-earth beliefs are often shaped and sustained by a worldview that combines religious elements with distrust in conventional sciences, reinforcing the belief in an alternative narrative grounded in literal interpretations of religious texts;

**Rejection of science as a central part of the narrative:** Flat-earth communities frequently engage in discussions about gravity and the evolution of species to reject established scientific concepts. This rejection not only questions the shape of the Earth but also challenges the entire structure of scientific knowledge, reinforcing a worldview where science is seen as a fallacious construct designed to deceive;

**Attraction to global catastrophe and apocalyptic theories:** Themes such as the biblical apocalypse and magnetic pole reversal are closely connected to flat-earth beliefs, suggesting that these communities often associate the belief in a flat Earth with catastrophic predictions and with dogmatic narratives stemming from spirituality. These apocalyptic narratives function as a logical extension of general distrust in science and authorities, offering an alternative explanation for global events;

**Persistence and resilience of flat-earth narratives:** Despite efforts to dismantle these beliefs, flat-earth narratives have shown remarkable resilience, maintaining active and cohesive discussions over the years. The analysis of time series shows that, although there have been peaks of interest, discussions have stabilized at a still-elevated level, indicating that these beliefs have become deeply rooted in the digital sphere;

**Flat-earth as a test of ideological commitment:** The interconnectivity between flat-earth beliefs and other conspiracy theories suggests that involvement with this narrative functions as a test of ideological commitment for its adherents. Once individuals engage with flat-earth beliefs, they are often exposed to a broader range of conspiracy narratives, deepening their engagement within the conspiratorial ecosystem and complicating the deconstruction of beliefs.

### 4.2. Future works

Based on the key findings of this study, several promising directions can be explored for future research. The intersection between flat-earth beliefs and other apocalyptic narratives suggests a fertile field for investigations aimed at understanding how these conspiracy theories are interconnected and mutually reinforcing. For example, future studies could focus on how the narratives of the New World Order and Apocalypse and Survivalism function as bridges to flat-earth beliefs, fueling adherence to other extreme beliefs. Additionally, it would be relevant to investigate how global crises, such as the COVID-19 pandemic, catalyze the spread of these theories, offering new perspectives on the evolution and adaptation of flat-earth beliefs during periods of uncertainty.

Another critical point for future investigations is the centralization observed in flat-earth networks, where a small number of communities act as dissemination hubs. Research could focus on mapping these central structures to identify how they maintain cohesion and influence within the network. Understanding the role of these hubs could provide valuable insights for developing intervention strategies aimed at more effectively disrupting the spread of disinformation.

The relationship between flat-earth beliefs and religion also opens up a range of possibilities for future studies. Considering that these communities use biblical interpretations to legitimize the rejection of science, it is essential to explore how these narratives are sustained and perpetuated. Investigations that address the psychology behind the fusion of religious beliefs and conspiracy theories could reveal the reasons why these ideas are so resistant to external interventions, facilitating the development of more targeted and effective correction campaigns.

Furthermore, the importance of digital platforms in the spread of flat-earth beliefs cannot be underestimated. Future studies could investigate how the architecture of these platforms facilitates the dissemination of disinformation and how flat-earth communities adapt their communication strategies in the face of restrictions or censorship in other media. Analyzing the migration of discussions to alternative platforms and the impact of this on the spread of radical ideas would also be a promising area for exploration.

Finally, the persistence of conspiracy narratives over time, even after repeated scientific debunking, indicates the need to understand the mechanisms that allow these beliefs to be reintroduced and gain traction again. Research could focus on how these cycles of resilience and reintroduction occur, particularly during critical moments, and how they can be effectively interrupted. Understanding these processes may be key to developing long-term strategies aimed at slowing the spread of disinformation and protecting public discourse from the harmful consequences of these beliefs.

## 6. Author biography

**Ergon Cugler de Moraes Silva** has a Master's degree in Public Administration and Government (FGV), Postgraduate MBA in Data Science & Analytics (USP) and Bachelor's degree in Public Policy Management (USP). He is associated with the Bureaucracy Studies Center (NEB FGV), collaborates with the Interdisciplinary Observatory of Public Policies (OIPP USP), with the Study Group on Technology and Innovations in Public Management (GETIP USP) with the Monitor of Political Debate in the Digital Environment (Monitor USP) and with the Working Group on Strategy, Data and Sovereignty of the Study and Research Group on International Security of the Institute of International Relations of the University of Brasília (GEPSI UnB). He is also a researcher at the Brazilian Institute of Information in Science and Technology (IBICT), where he works for the Federal Government on strategies against disinformation. Brasília, Federal District, Brazil. Web site: https://ergoncugler.com/.

[ versão português / english above ]

# **Comunidades de terraplanismo no Telegram brasileiro:** quando a fé é usada para para questionar a existência da gravidade enquanto fenômeno da física

*Ergon Cugler de Moraes Silva*

Instituto Brasileiro de Informação
em Ciência e Tecnologia (IBICT)
Brasília, Distrito Federal, Brasil

contato@ergoncugler.com
www.ergoncugler.com

**Resumo**

As teorias da conspiração relacionadas ao terraplanismo têm ganhado força no Telegram brasileiro, especialmente em momentos de crise global, como a Pandemia da COVID-19, onde a desconfiança nas instituições científicas e governamentais se intensificou. Dessa forma, esse estudo busca responder à pergunta de pesquisa: **como são caracterizadas e articuladas as comunidades de teorias da conspiração brasileiras sobre temáticas de terraplanismo no Telegram?** Vale ressaltar que este estudo faz parte de uma série de um total de sete estudos que possuem como objetivo principal compreender e caracterizar as comunidades brasileiras de teorias da conspiração no Telegram. Esta série de sete estudos está disponibilizada abertamente e originalmente no arXiv da Cornell University, aplicando um método espelhado nos sete estudos, mudando apenas o objeto temático de análise e provendo uma replicabilidade de investigação, inclusive com códigos próprios e autorais elaborados, somando-se à cultura de software livre e de código aberto. No que diz respeito aos principais achados deste estudo, observa-se: Durante a Pandemia, discussões terraplanistas aumentaram 400%, impulsionadas pela desconfiança nas instituições científicas; Comunidades terraplanistas atuam como portais para outras teorias conspiratórias, como a Nova Ordem Mundial; Apesar de menor, a rede terraplanista possui grupos influentes que disseminam conteúdo e perpetuam narrativas; Temas religiosos como “Deus” e “bíblia” são centrais nas discussões, combinando elementos religiosos com desconfiança na ciência; Comunidades terraplanistas utilizam temas como gravidade para desafiar conceitos científicos estabelecidos, reforçando uma visão alternativa do mundo.

## **Principais descobertas**

- ➔ Durante o auge da Pandemia da COVID-19, discussões em comunidades de terraplanismo no Telegram brasileiro aumentaram 400% entre 2019 e 2020, impulsionadas pela desconfiança nas instituições científicas e governamentais, refletindo como crises globais intensificam a disseminação de desinformação;

- ➔ Comunidades de terraplanismo não apenas atraem novos adeptos, mas também funcionam como portais para outras teorias conspiratórias, como a Nova Ordem Mundial e o Ocultismo, com 1.934 links vindos da NOM e 797 de Apocalipse e Sobrevivência, demonstrando a interconectividade dessas narrativas;

- ➔ A rede terraplanista é relativamente menor, se comparada com outra comunidades de teorias da conspiração, mas com poucos grupos no Telegram atuando como *hubs* principais na disseminação de conteúdo, detendo uma capacidade desproporcional de influenciar o discurso e perpetuar narrativas, enquanto grupos menores orbitam em torno deles;

➔ Discussões sobre a Nova Ordem Mundial são predominantes nas comunidades terraplanistas, com 724 links direcionados a essa temática, evidenciando que a narrativa de controle global é usada para legitimar e reforçar as crenças terraplanistas, criando uma sobreposição temática.

➔ A disseminação do terraplanismo depende fortemente de plataformas digitais, com termos como "canal" e "Telegram" sendo centrais nas discussões. A prevalência dessas plataformas, especialmente o Telegram, destaca a importância na manutenção e expansão de comunidades;

➔ Análises revelam uma forte presença de termos religiosos como "Deus" e "bíblia" nas discussões terraplanistas, sugerindo que essas crenças são moldadas por uma visão de mundo que combina elementos religiosos com desconfiança na ciência, sustentando alternativas;

➔ Comunidades terraplanistas frequentemente utilizam temas como gravidade e evolução das espécies para rejeitar conceitos científicos estabelecidos, desafiando não apenas a forma da Terra, mas toda estrutura do conhecimento científico, reforçando uma visão de mundo grupal;

➔ Temas como apocalipse bíblico e inversão de polos magnéticos são intimamente conectados ao terraplanismo, sugerindo que essas comunidades associam a crença na Terra plana a previsões catastróficas, reforçando narrativas apocalípticas que oferecem explicações alternativas para crises globais;

➔ Apesar dos esforços para desmantelar essas crenças, as narrativas terraplanistas mostram resiliência, mantendo discussões ativas e coesas ao longo dos anos, com níveis elevados de interesse que indicam um enraizamento profundo dessas crenças no ambiente digital;

➔ A interconectividade entre terraplanismo e outras teorias conspiratórias sugere que o envolvimento com essa narrativa funciona como um teste de comprometimento ideológico, expondo os indivíduos a uma gama mais ampla de conspirações e dificultando a desconstrução dessas crenças já inseridas enquanto sentimento de comunidade para com seus membros.

## 1. Introdução

Após percorrer milhares de comunidades brasileiras de teorias da conspiração no Telegram, extrair dezenas de milhões de conteúdos dessas comunidades, elaborados e/ou compartilhados por milhões de usuários que as compõem, este estudo tem o objetivo de compor uma série de um total de sete estudos que tratam sobre o fenômeno das teorias da conspiração no Telegram, adotando o Brasil como estudo de caso. Com as abordagens de identificação implementadas, foi possível alcançar um total de 30 comunidades de teorias da conspiração brasileiras no Telegram sobre temáticas de terraplanismo, estas somando 632.901de conteúdos publicados entre junho de 2019 (primeiras publicações) até agosto de 2024 (realização deste estudo), com 38.520 usuários somados dentre as comunidades. Dessa forma, este estudo tem como objetivo compreender e caracterizar as comunidades sobre temáticas de terraplanismo presentes nessa rede brasileira de teorias da conspiração identificada no Telegram.

Para tal, será aplicado um método espelhado em todos os sete estudos, mudando apenas o objeto temático de análise e provendo uma replicabilidade de investigação. Assim, abordaremos técnicas para observar as conexões, séries temporais, conteúdos e sobreposições temáticas das comunidades de teorias da conspiração. Além desse estudo, é possível encontrar os seis demais disponibilizados abertamente e originalmente no arXiv da Cornell University. Essa série contou com a atenção redobrada para garantir a integridade dos dados e o respeito à privacidade dos usuários, conforme a legislação brasileira prevê (Lei nº 13.709/2018).

Portanto questiona-se: **como são caracterizadas e articuladas as comunidades de teorias da conspiração brasileiras sobre temáticas de terraplanismo no Telegram?**

## 2. Materiais e métodos

A metodologia deste estudo está organizada em três subseções, sendo: **2.1. Extração de dados**, que descreve o processo e as ferramentas utilizadas para coletar as informações das comunidades no Telegram; **2.2. Tratamento de dados**, onde são abordados os critérios e métodos aplicados para classificar e anonimizar os dados coletados; e **2.3. Abordagens para análise de dados**, que detalha as técnicas utilizadas para investigar as conexões, séries temporais, conteúdos e sobreposições temáticas das comunidades de teorias da conspiração.

### 2.1. Extração de dados

Este projeto teve início em fevereiro de 2023, com a publicação da primeira versão do TelegramScrap (Silva, 2023), uma ferramenta própria e autoral, de software livre e de código aberto, que faz uso da Application Programming Interface (API) da plataforma Telegram por meio da biblioteca Telethon e organiza ciclos de extração de dados de grupos e canais abertos no Telegram. Ao longo dos meses, a base de dados pôde ser ampliada e qualificada fazendo uso de quatro abordagens de identificação de comunidades de teorias da conspiração:

**(i) Uso de palavras chave:** no início do projeto, foram elencadas palavras-chave para identificação diretamente no buscador de grupos e canais brasileiros no Telegram, tais como "apocalipse", "sobrevivencialismo", "mudanças climáticas", "terra plana", "teoria da conspiração", "globalismo", "nova ordem mundial", "ocultismo", "esoterismo", "curas alternativas", "qAnon", "reptilianos", "revisionismo", "alienígenas", dentre outras. Essa primeira abordagem forneceu algumas comunidades cujos títulos e/ou descrições dos grupos e canais contassem com os termos explícitos relacionados a teorias da conspiração. Contudo, com o tempo foi possível identificar outras diversas comunidades cujas palavras-chave elencadas não davam conta de abarcar, algumas inclusive propositalmente com caracteres trocados para dificultar a busca de quem a quisesse encontrar na rede;

**(ii) Mecanismo de recomendação de canais do Telegram:** com o tempo, foi identificado que canais do Telegram (exceto grupos) contam com uma aba de recomendação chamada de "canais similares", onde o próprio Telegram sugere dez canais que tenham alguma similaridade com o canal que se está observando. A partir desse mecanismo de recomendação do próprio Telegram, foi possível encontrar mais comunidades de teorias da conspiração brasileiras, com estas sendo recomendadas pela própria plataforma;

**(iii) Abordagem de bola de neve para identificação de convites:** após algumas comunidades iniciais serem acumuladas para a extração, foi elaborado um algoritmo próprio autoral de identificação de urls que contivessem "t.me/", sendo o prefixo de qualquer convite para grupos e canais do Telegram. Acumulando uma frequência de centenas de milhares de links que atendessem a esse critério, foram elencados os endereços únicos e contabilizadas as suas repetições. Dessa forma, foi possível fazer uma investigação de novos grupos e canais brasileiros mencionados nas próprias mensagens dos já investigados, ampliando a rede. Esse processo foi sendo repetido periodicamente, buscando identificar novas comunidades que tivessem identificação com as temáticas de teorias da conspiração no Telegram;

**(iv) Ampliação para tweets publicados no X que mencionassem convites:** com o objetivo de diversificar ainda mais a fonte de comunidades de teorias da conspiração brasileiras no Telegram, foi elaborada uma query de busca própria que pudesse identificar as palavras-chave de temáticas de teorias da conspiração, porém usando como fonte tweets publicados no X (antigo Twitter) e que, além de conter alguma das palavras-chave, contivesse também o "t.me/", sendo o prefixo de qualquer convite para grupos e canais do Telegram, "https://x.com/search?q=lang%3Apt%20%22t.me%2F%22%20TERMO-DE-BUSCA".

Com as abordagens de identificação de comunidades de teorias da conspiração implementadas ao longo de meses de investigação e aprimoramento de método, foi possível construir uma base de dados do projeto com um total de 855 comunidades de teorias da conspiração brasileiras no Telegram (considerando as demais temáticas também não incluídas nesse estudo), estas somando 27.227.525 de conteúdos publicados entre maio de 2016 (primeiras publicações) até agosto de 2024 (realização deste estudo), com 2.290.621 usuários somados dentre as comunidades brasileiras. Há de se considerar que este volume de usuários conta com dois elementos, o primeiro é que trata-se de uma variável, pois usuários podem entrar e sair diariamente, portanto este valor representa o registrado na data de extração de publicações da comunidade; além disso, é possível que um mesmo usuário esteja em mais de um grupo e, portanto, é contabilizado mais de uma vez. Nesse sentido, o volume ainda sinaliza ser expressivo, mas pode ser levemente menor quando considerado o volume de cidadãos deduplicados dentro dessas comunidades brasileiras de teorias da conspiração.

### 2.2. Tratamento de dados

Com todos os grupos e canais brasileiros de teorias da conspiração no Telegram extraídos, foi realizada uma classificação manual considerando o título e a descrição da comunidade. Caso houvesse menção explícita no título ou na descrição da comunidade a alguma temática, esta foi classificada entre: (i) "Anticiência"; (ii) "Anti-Woke e Gênero"; (iii) "Antivax"; (iv) "Apocalipse e Sobrevivencialismo"; (v) "Mudanças Climáticas"; (vi) Terra Plana; (vii) "Globalismo"; (viii) "Nova Ordem Mundial"; (ix) "Ocultismo e Esoterismo"; (x) "Off Label e Charlatanismo"; (xi) "QAnon"; (xii) "Reptilianos e Criaturas"; (xiii) "Revisionismo e Discurso de Ódio"; (xiv) "OVNI e Universo". Caso não houvesse nenhuma menção explícita relacionada às temáticas no título ou na descrição da comunidade, esta foi classificada como (xv) "Conspiração Geral". No Quadro a seguir, podemos observar as métricas relacionadas à classificação dessas comunidades de teorias da conspiração no Brasil.

**Tabela 01.** Comunidades de teorias da conspiração no Brasil (métricas até agosto de 2024)

| Categorias | Grupos | Usuários | Publicações | Comentários | Total |
|---|---|---|---|---|---|
| **Anticiência** | 22 | 58.138 | 187.585 | 784.331 | 971.916 |
| **Anti-*Woke* e Gênero** | 43 | 154.391 | 276.018 | 1.017.412 | 1.293.430 |
| **Antivacinas (*Antivax*)** | 111 | 239.309 | 1.778.587 | 1.965.381 | 3.743.968 |
| **Apocalipse e Sobrevivência** | 33 | 109.266 | 915.584 | 429.476 | 1.345.060 |
| **Mudanças Climáticas** | 14 | 20.114 | 269.203 | 46.819 | 316.022 |
| **Terraplanismo** | 33 | 38.563 | 354.200 | 1.025.039 | 1.379.239 |
| **Conspirações Gerais** | 127 | 498.190 | 2.671.440 | 3.498.492 | 6.169.932 |
| **Globalismo** | 41 | 326.596 | 768.176 | 537.087 | 1.305.263 |
| **Nova Ordem Mundial (NOM)** | 148 | 329.304 | 2.411.003 | 1.077.683 | 3.488.686 |
| **Ocultismo e Esoterismo** | 39 | 82.872 | 927.708 | 2.098.357 | 3.026.065 |
| **Medicamentos *off label*** | 84 | 201.342 | 929.156 | 733.638 | 1.662.794 |
| **QAnon** | 28 | 62.346 | 531.678 | 219.742 | 751.420 |
| **Reptilianos e Criaturas** | 19 | 82.290 | 96.262 | 62.342 | 158.604 |
| **Revisionismo e Ódio** | 66 | 34.380 | 204.453 | 142.266 | 346.719 |
| **OVNI e Universo** | 47 | 58.912 | 862.358 | 406.049 | 1.268.407 |
| **Total** | **855** | **2.296.013** | **13.183.411** | **14.044.114** | **27.227.525** |

Fonte: Elaboração própria (2024).

Com esse volume de dados extraídos, foi possível segmentar para apresentar neste estudo apenas comunidades e conteúdos referentes às temáticas de terraplanismo. Em paralelo, as demais temáticas de comunidades brasileiras de teorias da conspiração também contaram com estudos elaborados para a caracterização da extensão e da dinâmica da rede, estes sendo disponibilizados abertamente e originalmente no arXiv da Cornell University.

Além disso, cabe citar que apenas foram extraídas comunidades abertas, isto é, não apenas identificáveis publicamente, mas também sem necessidade de solicitação para acessar ao conteúdo, estando aberto para todo e qualquer usuário com alguma conta do Telegram sem que este necessite ingressar no grupo ou canal. Além disso, em respeito à legislação brasileira e especialmente da Lei Geral de Proteção de Dados Pessoais (LGPD), ou Lei nº 13.709/2018, que trata do controle da privacidade e do uso/tratamento de dados pessoais, todos os dados extraídos foram anonimizados para a realização de análises e investigações. Dessa forma, nem mesmo a identificação das comunidades é possível por meio deste estudo, estendendo aqui a privacidade do usuário ao anonimizar os seus dados para além da própria comunidade à qual ele se submeteu ao estar em um grupo ou canal público e aberto no Telegram.

### 2.3. Abordagens para análise de dados

Totalizando 30 comunidades selecionadas nas temáticas de terraplanismo, contendo 632.901 publicações e 38.520 usuários somados, quatro abordagens serão utilizadas para investigar as comunidades de teorias da conspiração selecionadas para o escopo do estudo. Tais métricas são detalhadas no Quadro a seguir:

**Tabela 02.** Comunidades selecionadas para análise (métricas até agosto de 2024)

| Categoria | Grupos | Usuários | Publicações | Comentários | Total |
|---|---|---|---|---|---|
| **Terraplanismo** | 30 | 38.520 | 188.068 | 444.833 | 632.901 |

Fonte: Elaboração própria (2024).

**(i) Rede:** com a elaboração de um algoritmo próprio para a identificação de mensagens que contenham o termo de "t.me/" (de convite para entrarem em outras comunidades), propomos apresentar volumes e conexões observadas sobre como **(a)** as comunidades da temática de terraplanismo circulam convites para que os seus usuários conheçam mais grupos e canais da mesma temática, reforçando os sistemas de crença que comungam; e como **(b)** essas mesmas comunidades circulam convites para que os seus usuários conheçam grupos e canais que tratem de outras teorias da conspiração, distintas de seu propósito explícito. Esta abordagem é interessante para observar se essas comunidades utilizam a si próprias como fonte de legitimação e referência e/ou se embasam-se em demais temáticas de teorias da conspiração, inclusive abrindo portas para que seus usuários conheçam outras conspirações. Além disso, cabe citar o estudo de Rocha *et al.* (2024) em que uma abordagem de identificação de rede também foi aplicada em comunidades do Telegram, porém observando conteúdos similares a partir de um ID gerado para cada mensagem única e suas similares;

**(ii) Séries temporais:** utiliza-se da biblioteca "Pandas" (McKinney, 2010) para organizar os data frames de investigação, observando **(a)** o volume de publicações ao longo dos meses; e **(b)** o volume de engajamento ao longo dos meses, considerando metadados de visualizações, reações e comentários coletados na extração; Além da volumetria, a biblioteca "Plotly" (Plotly Technologies Inc., 2015) viabilizou a representação gráfica dessa variação;

**(iii) Análise de conteúdo:** além da análise geral de palavras com identificação das frequências, são aplicadas séries temporais na variação das palavras mais frequentes ao longo dos semestres — observando entre junho de 2019 (primeiras publicações) até agosto de 2024 (realização deste estudo). E com as bibliotecas "Pandas" (McKinney, 2010) e "WordCloud" (Mueller, 2020), os resultados são apresentados tanto volumetricamente quanto graficamente;

**(iv) Sobreposição de agenda temática:** seguindo a abordagem proposta por Silva & Sátiro (2024) para identificação de sobreposição de agenda temática em comunidades do Telegram, utilizamos o modelo "BERTopic" (Grootendorst, 2020). O BERTopic é um algoritmo de modelagem de tópicos que facilita a geração de representações temáticas a partir de grandes quantidades de textos. Primeiramente, o algoritmo extrai embeddings dos documentos usando modelos transformadores de sentenças, como o "all-MiniLM-L6-v2". Em seguida, essas embeddings têm sua dimensionalidade reduzida por técnicas como "UMAP", facilitando o processo de agrupamento. A clusterização é realizada usando "HDBSCAN", uma técnica baseada em densidade que identifica clusters de diferentes formas e tamanhos, além de detectar outliers. Posteriormente, os documentos são tokenizados e representados em uma estrutura de bag-of-words, que é normalizada (L1) para considerar as diferenças de tamanho entre os clusters. A representação dos tópicos é refinada usando uma versão modificada do "TF-IDF", chamada "Class-TF-IDF", que considera a importância das palavras dentro de cada cluster (Grootendorst, 2020). Cabe destacar que, antes de aplicar o BERTopic, realizamos a limpeza da base removendo "stopwords" em português, por meio da biblioteca "NLTK" (Loper & Bird, 2002). Para a modelagem de tópicos, utilizamos o backend "loky" para otimizar o desempenho durante o ajuste e a transformação dos dados.

Em síntese, a metodologia aplicada compreendeu desde a extração de dados com a ferramenta própria autoral TelegramScrap (Silva, 2023), até o tratamento e a análise de dados coletados, utilizando diversas abordagens para identificar e classificar comunidades de teorias da conspiração brasileiras no Telegram. Cada uma das etapas foi cuidadosamente implementada para garantir a integridade dos dados e o respeito à privacidade dos usuários, conforme a legislação brasileira prevê. A seguir, serão apresentados os resultados desses dados, com o intuito de revelar as dinâmicas e as características das comunidades estudadas.

## 3. Resultados

A seguir, os resultados são detalhados na ordem prevista na metodologia, iniciando com a caracterização da rede e suas fontes de legitimação e referência, avançando para as séries temporais, incorporando a análise de conteúdo e concluindo com a identificação de sobreposição de agenda temática dentre as comunidades de teorias da conspiração.

### 3.1. Rede

A análise da rede de comunidades de terraplanismo nos proporciona uma visão detalhada sobre como essas teorias são disseminadas e interconectadas dentro do ecossistema conspiratório. As figuras apresentadas exploram diferentes aspectos dessas conexões, revelando tanto a estrutura interna das redes de terraplanismo quanto seu papel como porta de entrada e saída para outras teorias conspiratórias. Ao analisar a rede interna (Figura 01), podemos observar que a disseminação do terraplanismo é amplamente centralizada, com algumas comunidades exercendo um papel de *hubs* centrais, onde se concentram as interações e compartilhamentos de conteúdo. Essa centralização sugere que um pequeno número de grupos tem grande influência na perpetuação dessas narrativas, enquanto comunidades menores, embora periféricas, ainda são impactadas pelo conteúdo centralizado.

Nas Figuras 02 e 03, que abordam as redes de portas de entrada e saída, fica evidente que o terraplanismo está conectado a outras teorias conspiratórias, como a Nova Ordem

Mundial e o Ocultismo. Isso indica que o terraplanismo pode tanto introduzir novos adeptos a um universo conspiratório mais amplo quanto ser o ponto de partida para que indivíduos migrem para discussões mais complexas e diversificadas. Essa interconectividade é essencial para entender como as crenças em torno da Terra plana podem facilitar a transição para outras teorias, ampliando o engajamento dos indivíduos dentro dessas comunidades.

Por fim, o fluxo de links de convites entre comunidades de terraplanismo (Figura 04) revela a dinâmica de recrutamento e radicalização que ocorre dentro dessas redes. O terraplanismo, ao ser vinculado a temas como Nova Ordem Mundial e Apocalipse, funciona como um gateway para narrativas mais profundas e ideologicamente carregadas. Essa figura ilustra como os adeptos do terraplanismo são direcionados para outras teorias conspiratórias centrais, reforçando o papel dessa teoria como um teste de comprometimento ideológico que, uma vez superado, leva os indivíduos a um envolvimento mais profundo com as teorias de grande escala. Assim, o terraplanismo não apenas mantém sua base de seguidores, mas também desempenha um papel crucial na integração de novos membros em uma rede mais ampla e interconectada de crenças conspiratórias.

**Figura 01.** Rede interna entre comunidades de terraplanismo

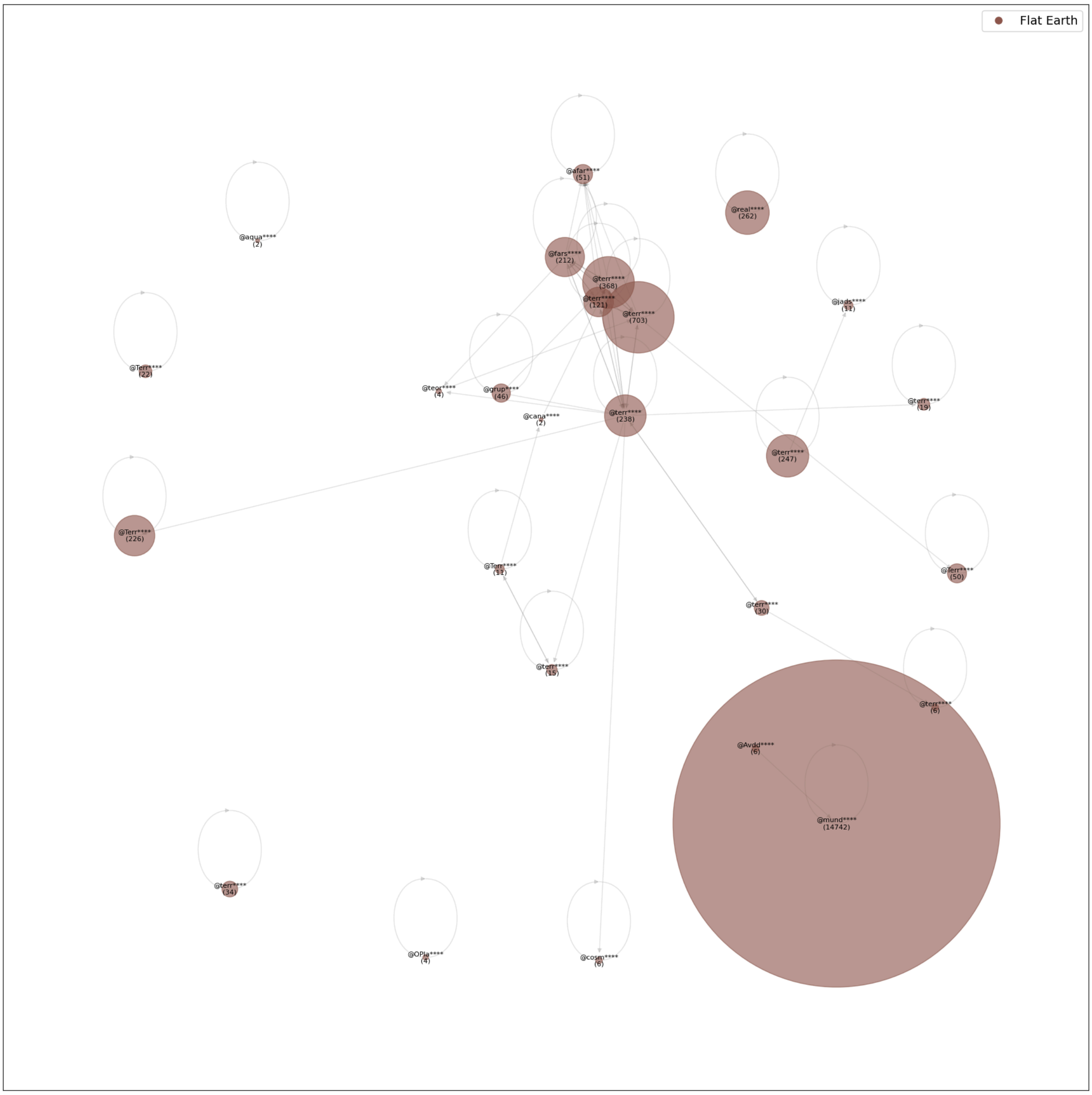

Fonte: Elaboração própria (2024).

A figura ilustra uma rede interna entre comunidades que promovem teorias de terraplanismo. As conexões revelam uma estrutura onde poucas comunidades atuam como *hubs* centrais, sendo os maiores nós indicativos das principais fontes de disseminação dessa teoria. A rede, embora menos densa comparada a outras, sugere uma centralização na propagação de informações, com uma forte interligação entre os grupos maiores e os menores que orbitam em torno deles. Esses *hubs* centrais parecem concentrar a maior parte das interações e compartilhamentos de conteúdo, o que indica que uma quantidade limitada de grupos exerce grande influência na perpetuação das narrativas terraplanistas. A presença de comunidades periféricas mais isoladas também sugere que, embora algumas subcomunidades possam não ter uma conexão direta com o núcleo central, elas ainda são impactadas pelo conteúdo disseminado pelos maiores influenciadores da rede.

**Figura 02.** Rede de comunidades que abrem portas para a temática (porta de entrada)

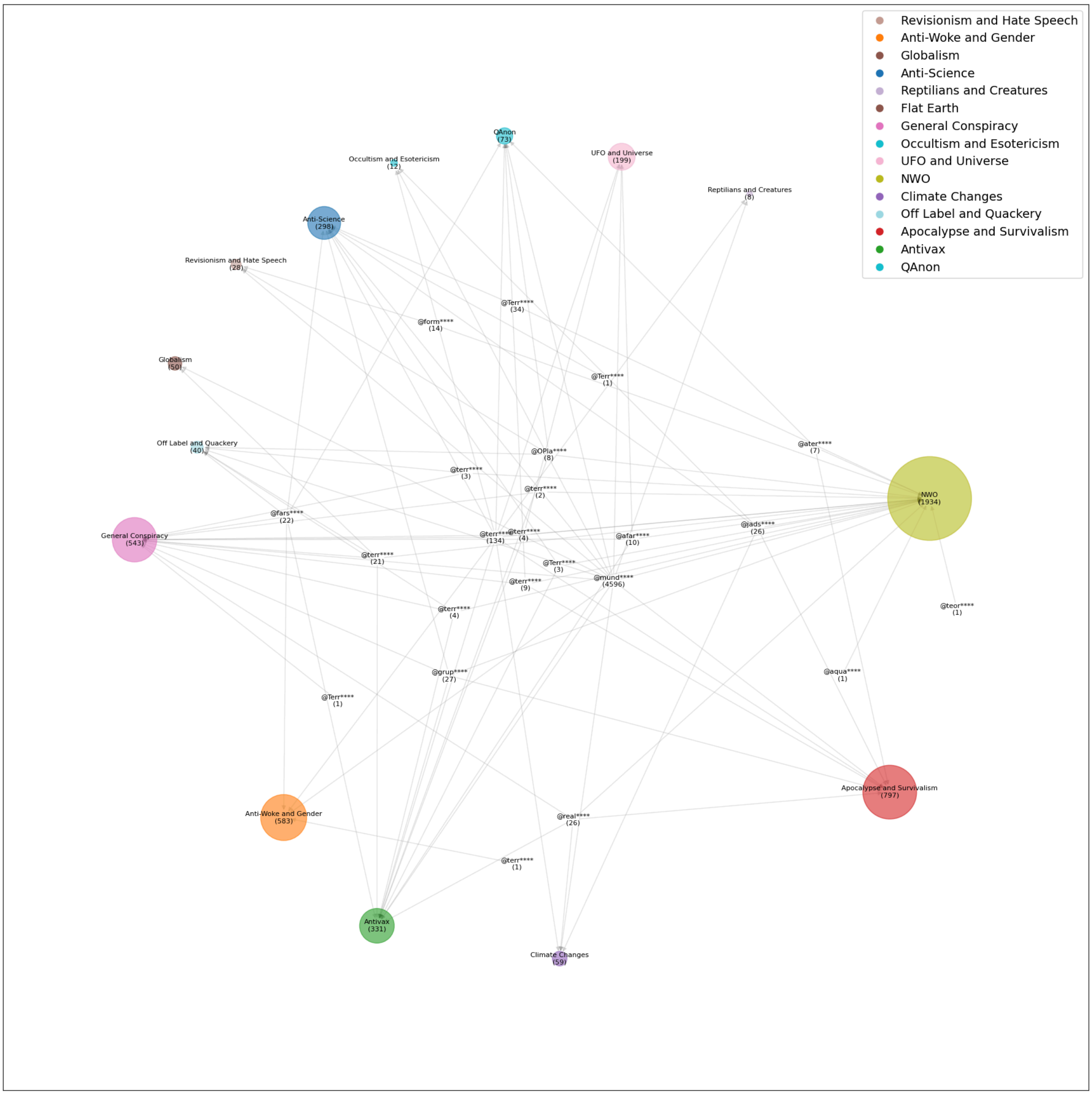

Fonte: Elaboração própria (2024).

A figura ilustra a rede de comunidades que servem como porta de entrada para o terraplanismo. Observa-se que, embora o terraplanismo seja uma teoria conspiratória com uma base relativamente isolada, ele ainda está conectado a outras comunidades que discutem temas similares, como a Nova Ordem Mundial e o Ocultismo. Essas conexões indicam que o terraplanismo pode ser uma porta de entrada para outras teorias mais amplas ou vice-versa. O gráfico sugere que, uma vez que indivíduos são expostos ao terraplanismo, há uma probabilidade de migração para outras teorias conspiratórias, com as comunidades maiores atuando como *hubs* que centralizam e disseminam conteúdos relacionados, facilitando a transição para outras áreas conspiratórias.

**Figura 03.** Rede de comunidades cuja temática abre portas (porta de saída)

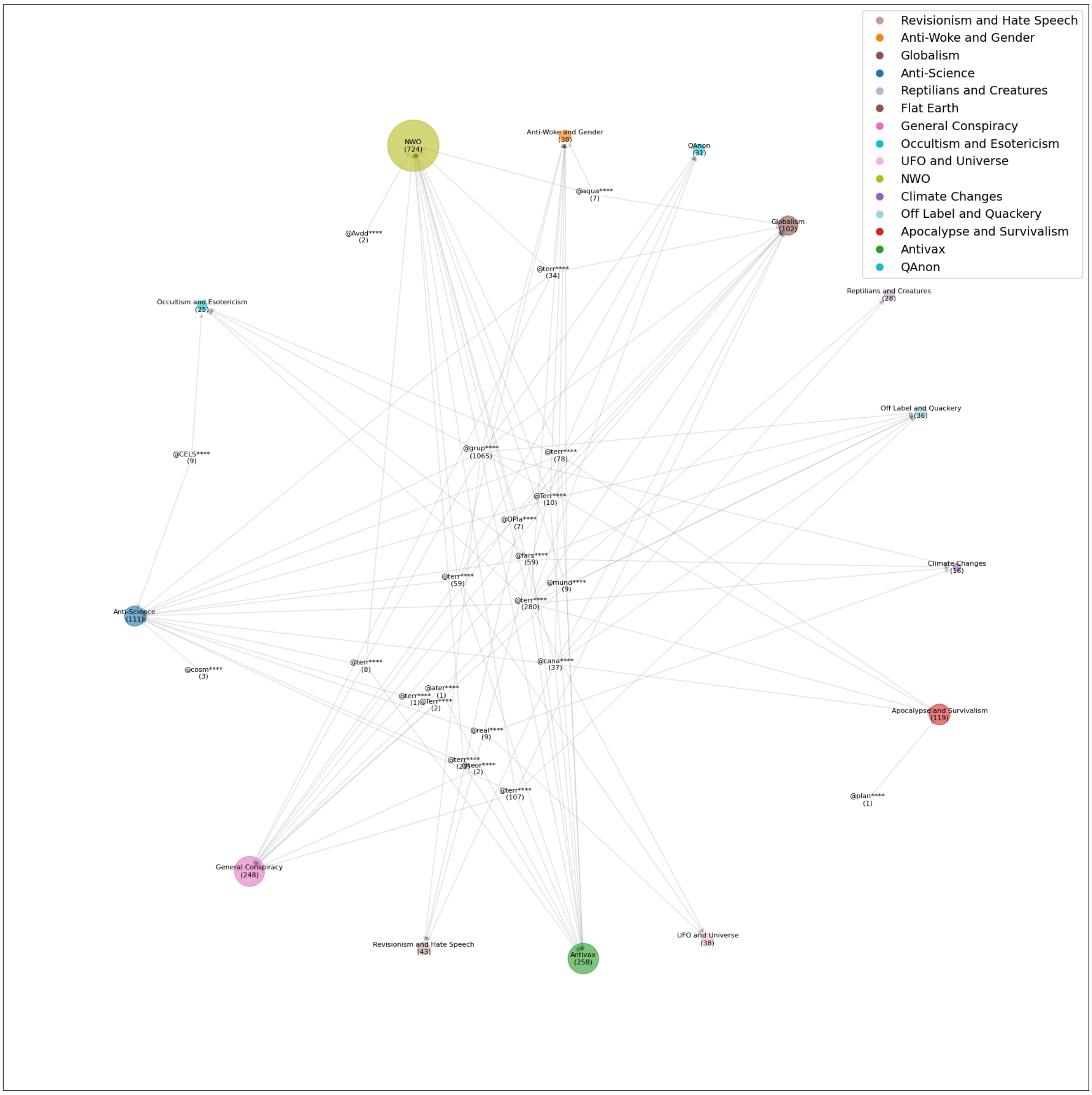

Fonte: Elaboração própria (2024).

Este gráfico revela as interconexões entre comunidades centradas na teoria da Terra plana e como essas comunidades podem agir como catalisadoras para a exploração de outras teorias da conspiração. Observa-se que, embora a teoria da Terra plana seja um nicho específico dentro do universo conspiratório, as comunidades associadas a ela têm conexões com uma gama diversificada de outras temáticas. A presença de grandes nós, como o da "Nova Ordem Mundial" e "Ocultismo e Esoterismo", sugere que seguidores da teoria da Terra plana podem facilmente transitar para discussões mais amplas e complexas, expandindo seu envolvimento dentro do ecossistema conspiratório. A figura ilustra como a teoria da Terra plana pode servir como uma "porta de saída" para outras teorias, mostrando que uma vez inseridos nesse nicho, os indivíduos são frequentemente expostos a um universo mais amplo de teorias correlatas, muitas vezes com uma visão distorcida e amplificada do mundo.

**Figura 04.** Fluxo de links de convites entre comunidades de terraplanismo

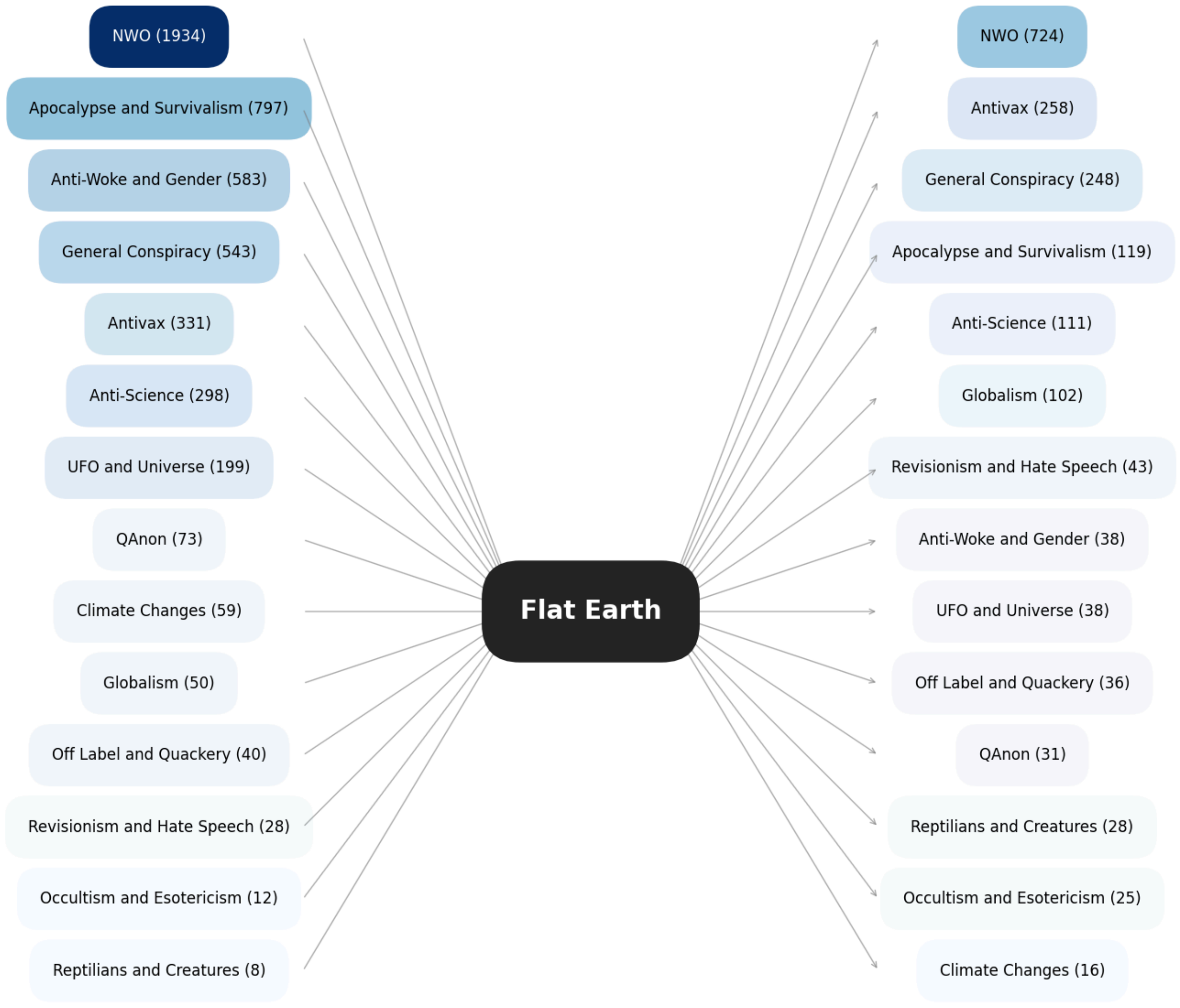

Fonte: Elaboração própria (2024).

O gráfico de fluxo de links de convites relacionados ao terraplanismo revela um cenário onde essa teoria, frequentemente tratada como um tema marginal e exótico, atua como uma porta de entrada significativa para um ecossistema mais amplo de teorias conspiratórias. A predominância de links provenientes da Nova Ordem Mundial (1.934) e de temas como Apocalipse e Sobrevivência (797) e Anti-*Woke* e Gênero (583) sugere que o terraplanismo serve como uma camada de acesso inicial a narrativas mais complexas e profundas. Essa dinâmica é crucial para a compreensão do papel que o terraplanismo desempenha no recrutamento e radicalização de indivíduos dentro dessas redes. O movimento "flat earth" não é apenas uma crença isolada; ele é um ponto de ancoragem que prepara os adeptos para abraçar narrativas mais amplas de resistência contra o conhecimento científico estabelecido e a ordem global. No sentido oposto, com a NOM recebendo 724 links de terraplanismo, observamos que os adeptos dessa teoria são rapidamente direcionados para narrativas mais centralizadas de controle global e resistência à ciência. Isso reflete a forma como o terraplanismo pode ser instrumentalizado para engajar indivíduos em discursos mais abrangentes e ideologicamente carregados, indicando que essa teoria funciona como uma espécie de teste de comprometimento ideológico que, uma vez superado, leva os indivíduos a um envolvimento mais profundo com teorias de grande escala.

### 3.2. Séries temporais

A análise das séries temporais nos permite observar como o terraplanismo, uma teoria conspiratória que desafia fundamentos científicos estabelecidos, ganhou tração em momentos-chave nos últimos anos. No gráfico a seguir, vemos como as menções a essa teoria cresceram abruptamente em 2020, possivelmente alimentadas pelas incertezas geradas pela Pandemia da COVID-19 e as subsequentes restrições globais. Este aumento foi seguido por um segundo pico em janeiro de 2021, coincidindo com um período de alta desconfiança política nos Estados Unidos. Estes picos revelam como crises globais e políticas podem catalisar a disseminação de teorias conspiratórias, criando um ambiente fértil para a proliferação de ideias radicais. À medida que avançamos no tempo, uma estabilização é notada, indicando uma possível normalização dessas discussões na cultura digital.

**Figura 05.** Gráfico de linhas do período

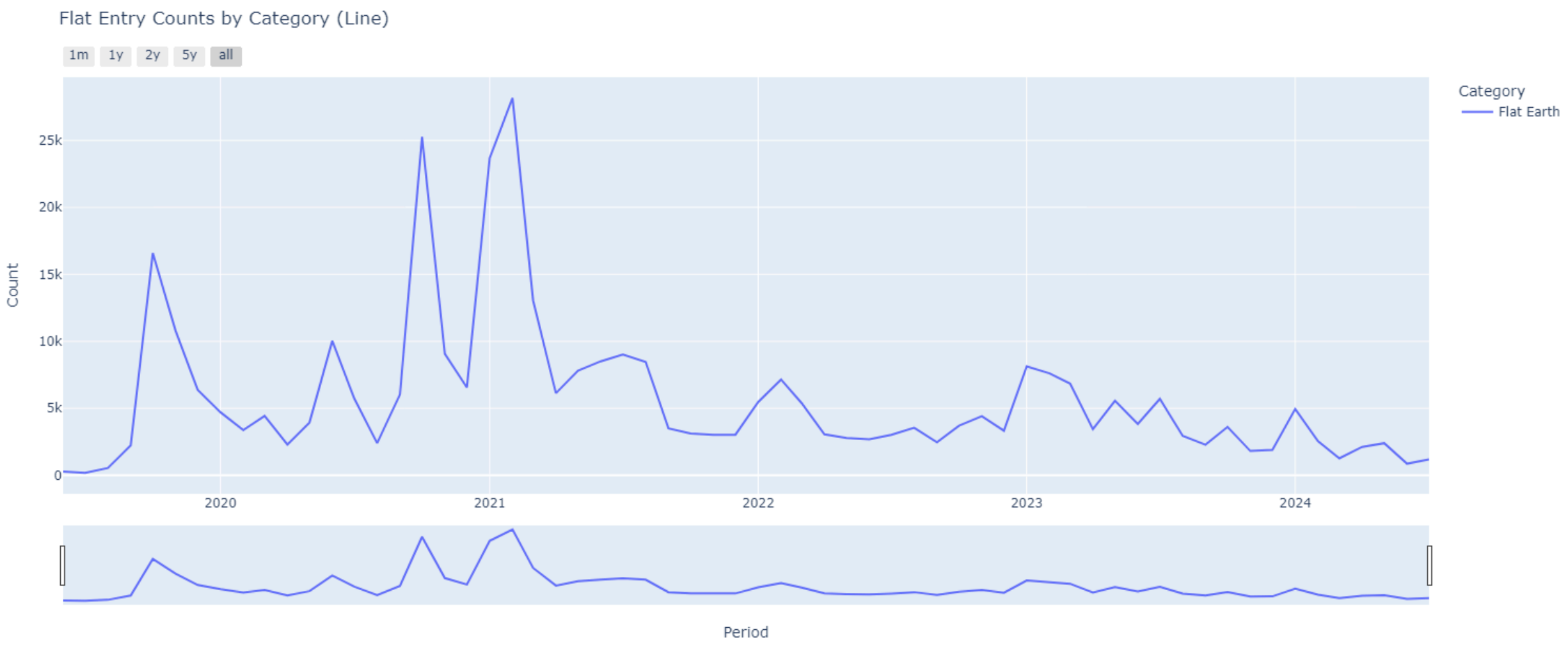

Fonte: Elaboração própria (2024).

No gráfico de terraplanismo, observamos que o primeiro pico significativo ocorre em meados de 2020, onde as menções aumentam de cerca de 500 para aproximadamente 2.500, o que representa um crescimento de 400% em um curto período. Este aumento coincide com o início das restrições de mobilidade global devido à COVID-19, quando teorias da conspiração sobre a "manipulação" pela ciência começaram a proliferar. Outro pico relevante ocorre em janeiro de 2021, onde as menções atingem novamente cerca de 2.500. Comparado ao volume inicial de 2020, isso representa uma variação de cerca de 300%. Este pico pode ser associado ao clima de desconfiança gerado pelas eleições presidenciais nos Estados Unidos, onde teorias da conspiração de diversos tipos, incluindo terraplanismo, se entrelaçaram com discussões políticas. Após esses picos, a partir de 2022, vemos uma estabilização em torno de 1.000 a 1.500 menções mensais, uma queda de aproximadamente 40% em relação aos picos anteriores. Isso sugere uma redução na intensidade das discussões, apesar do engajamento.

### 3.3. Análise de conteúdo

A análise de conteúdo das comunidades de terraplanismo através de nuvens de palavras proporciona uma visão detalhada sobre como certos termos e conceitos se consolidam e evoluem dentro dessas discussões ao longo do tempo. A partir das palavras mais frequentes, como "terra", "plana", "verdade", e "grupo", emergem as bases centrais do discurso terraplanista, revelando um padrão de linguagem que reflete tanto a tentativa de reafirmação de crenças conspiratórias quanto a resistência à ciência convencional. Essa análise permite observar como o terraplanismo não apenas questiona a forma da Terra, mas também está intimamente ligado a uma visão de mundo mais ampla que inclui elementos religiosos, como indicado pela presença do termo "Deus", e elementos de desconfiança em relação a instituições e autoridades. A ênfase em palavras como "canal" e "Telegram" ao longo dos anos sugere um foco na disseminação dessas ideias por meio de plataformas digitais, demonstrando a importância dessas mídias na perpetuação e expansão das crenças terraplanistas. Assim, a nuvem de palavras não só evidencia os principais temas discutidos, mas também revela a maneira como essas comunidades se organizam e se comunicam para sustentar e difundir suas convicções.

**Figura 06.** Nuvem de palavras consolidadas de terraplanismo

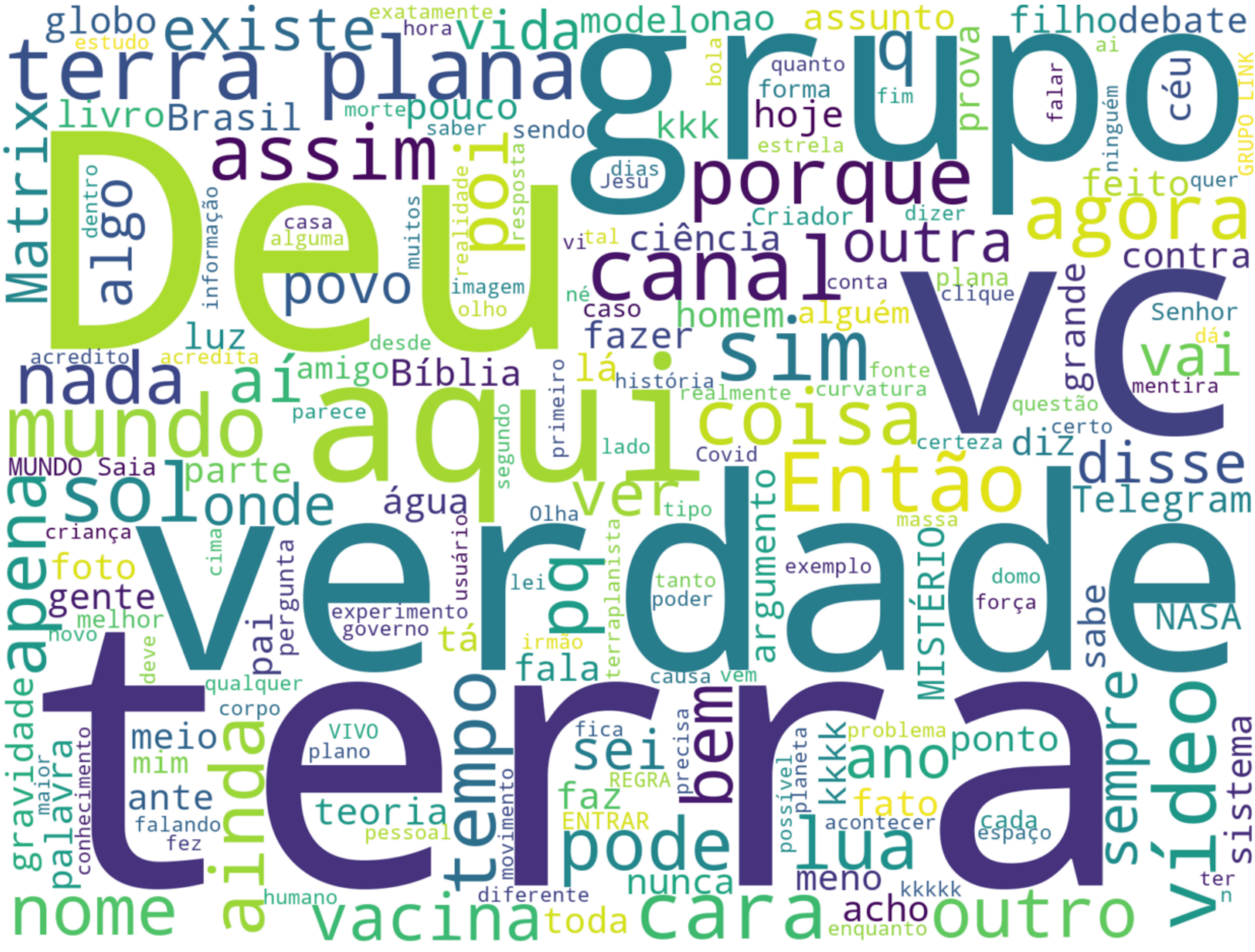

Fonte: Elaboração própria (2024).

A nuvem de palavras consolidada do terraplanismo revela a centralidade de termos como “terra”, “plana”, “verdade”, “grupo” e “agora”. O destaque dado à palavra “verdade” sugere um esforço contínuo por parte dos membros dessas comunidades em afirmar uma realidade alternativa, onde as afirmações sobre a forma da Terra são tomadas como uma verdade absoluta que se contrapõe ao conhecimento científico convencional. A palavra “grupo” evidencia o papel crucial das comunidades na disseminação dessas crenças, funcionando como espaços de reafirmação mútua e resistência contra a percepção dominante. Além disso, a presença de termos como “Deus”, “vida”, e “mundo” indica que a discussão dentro dessas comunidades não se limita a questões científicas, mas se estende a aspectos espirituais e existenciais, sugerindo que o terraplanismo é parte de uma visão de mundo mais ampla que combina elementos religiosos e conspiratórios. A recorrência da palavra “canal” aponta para a importância das mídias digitais, como YouTube e Telegram, na propagação dessas ideias, onde vídeos e grupos específicos desempenham um papel central na manutenção e expansão da comunidade.

**Quadro 01.** Nuvem de palavras em série temporal de terraplanismo

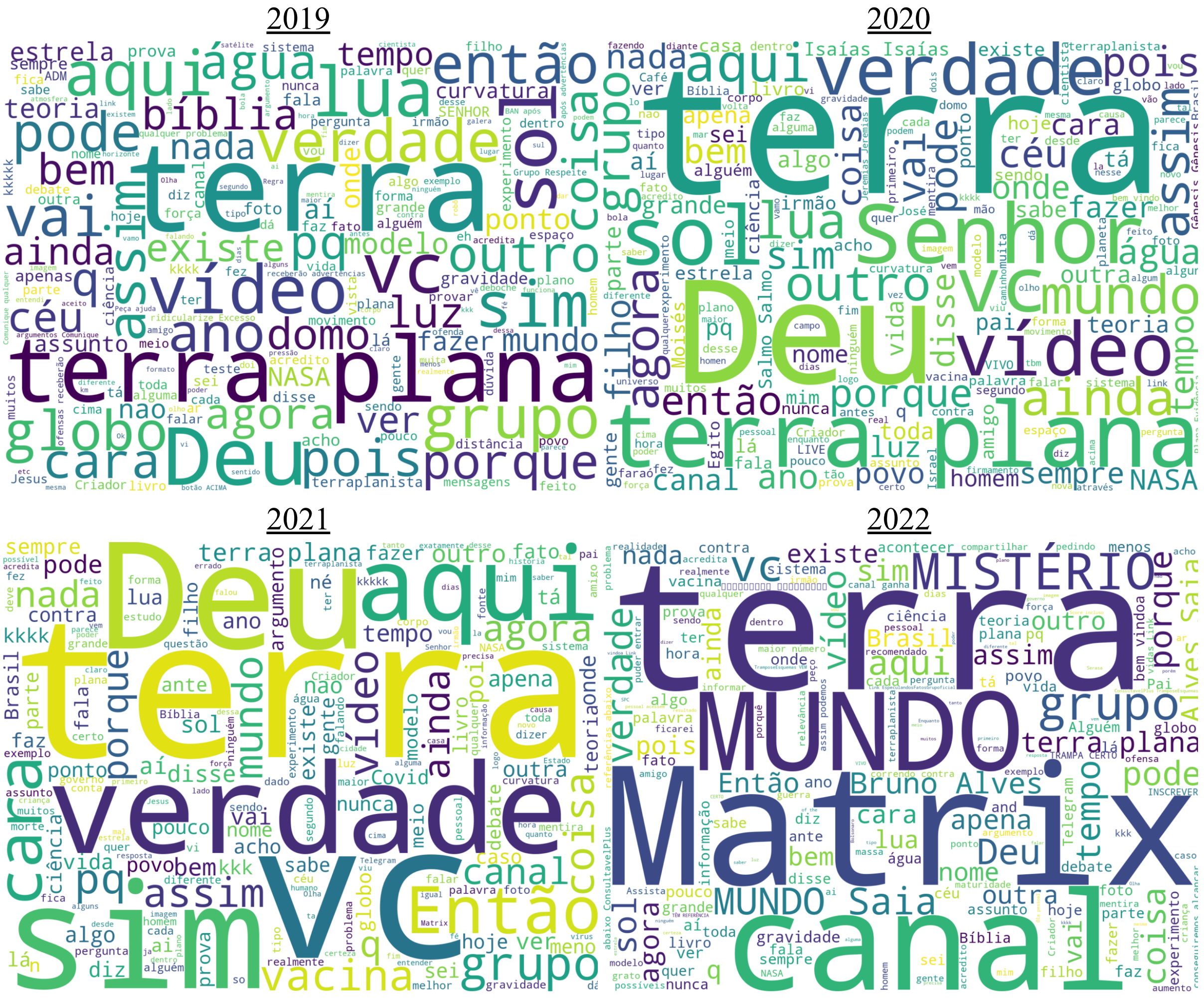

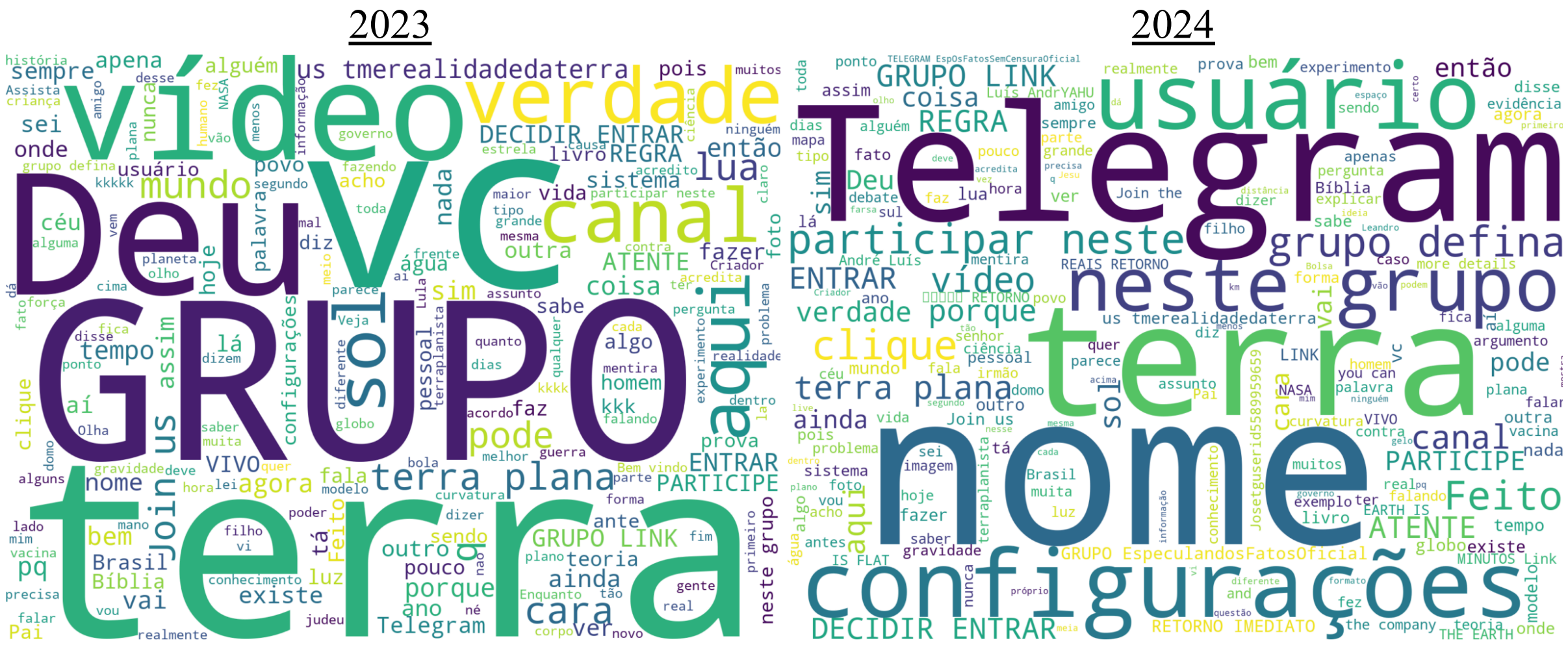

Fonte: Elaboração própria (2024).

O quadro de nuvens de palavras em série temporal para o terraplanismo revela como as discussões e prioridades dessas comunidades evoluíram ao longo dos anos. Em 2019, as palavras “terra” e “plana” já dominavam as discussões, indicando uma fase de afirmação dessas ideias básicas dentro da comunidade. A presença de termos como “lua” e “energia” também sugere um interesse em teorias que conectam o terraplanismo a conceitos mais esotéricos. Em 2020, com o impacto da Pandemia da COVID-19, observa-se a entrada de termos como “Matrix” e “mundo”, o que pode refletir um aumento nas discussões sobre controle e realidade simulada, frequentemente vinculadas à ideia de que a Pandemia faz parte de um plano maior de manipulação global. Em 2021, “verdade” e “canal” ganham destaque, sugerindo uma intensificação na disseminação de conteúdos que buscam descreditar a ciência oficial e promover narrativas alternativas através de canais de mídia digital. Nos anos seguintes, especialmente em 2023 e 2024, há uma maior diversificação de temas, com a inclusão de termos como “Telegram” e “configurações”, o que pode indicar uma adaptação das estratégias de comunicação, talvez em resposta a censuras ou restrições em outras plataformas. Essa evolução destaca como as comunidades terraplanistas não apenas persistem, mas também se adaptam ao contexto social e tecnológico, garantindo a expansão narrativa.

### 3.4. Sobreposição de agenda temática

As figuras a seguir representam uma análise de sobreposição de agenda temática em comunidades de teoria da conspiração, com um foco particular no terraplanismo. Através da visualização dos tópicos abordados por essas comunidades, é possível identificar como diferentes narrativas de desinformação se interligam, criando um discurso coeso que reforça as crenças centrais dessas comunidades. Esta análise busca destacar as conexões entre as teorias da conspiração e como temas aparentemente distintos, como ciência, religião e geopolítica, são utilizados para sustentar a narrativa do terraplanismo.

**Figura 07.** Temáticas de terraplanismo e rejeição à gravidade baseado na bíblia

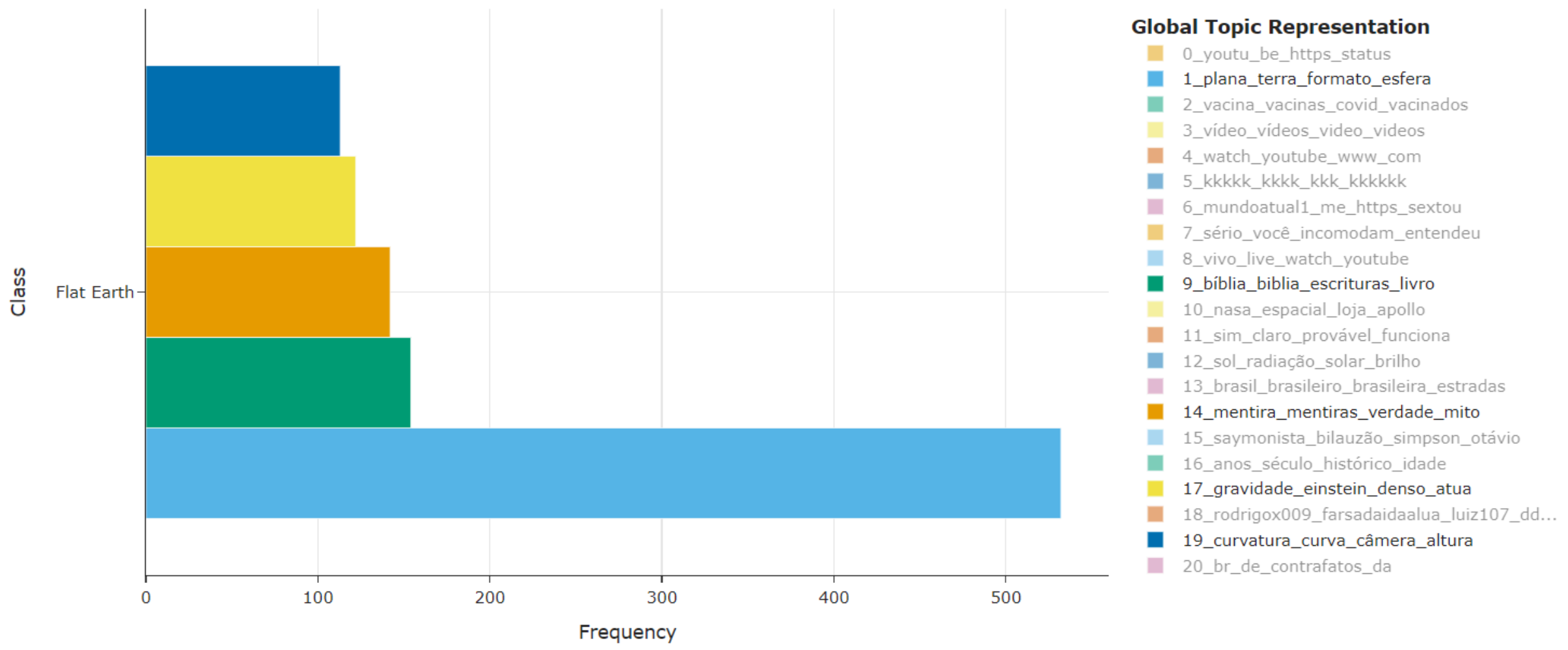

Fonte: Elaboração própria (2024).

A Figura 07 explora a relação entre o terraplanismo e a rejeição à gravidade, com base em interpretações bíblicas. A predominância de tópicos como "bíblia", "escrituras", e "livro" indica que essas comunidades utilizam passagens religiosas para justificar suas crenças de que a gravidade é um conceito inventado para enganar a humanidade. A sobreposição de tópicos relacionados à religião com a teoria da Terra plana sugere uma tentativa de legitimar essa crença através de uma leitura literal da bíblia, criando uma conexão entre ciência e fé que resiste a argumentos racionais baseados em evidências.

**Figura 08.** Temáticas de fé e demônios atrelados à temática de terraplanismo

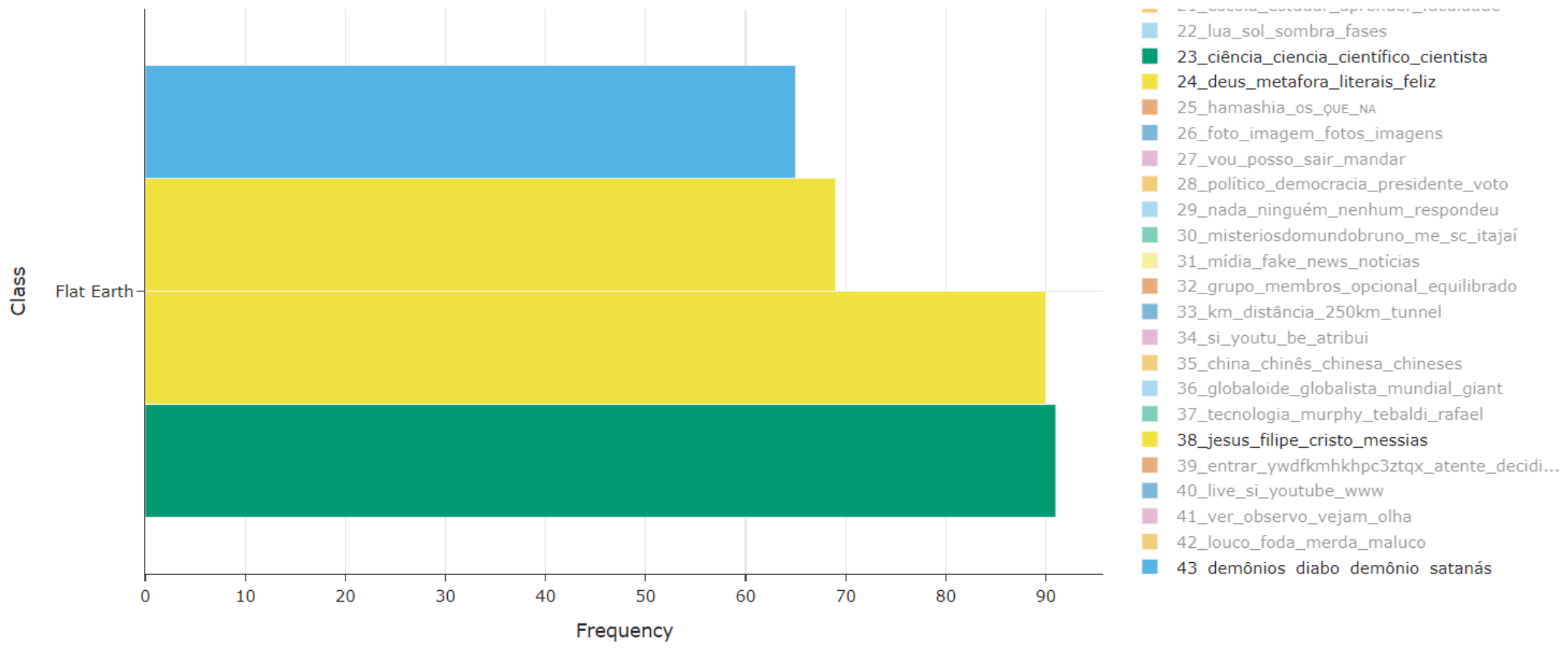

Fonte: Elaboração própria (2024).

Na Figura 08, é evidenciada a conexão entre o terraplanismo e temas de fé e demônios. Tópicos como "demônios", "diabo", e "satanás" indicam que as discussões nessas comunidades frequentemente associam a crença na Terra plana com uma batalha espiritual, onde a aceitação da ciência moderna é vista como um ato demoníaco. Essa abordagem fortalece a ideia de que a crença na Terra plana é parte de uma resistência espiritual contra forças malignas, consolidando o terraplanismo como uma crença que transcende o debate científico para se tornar uma questão moral e religiosa.

**Figura 09.** Temáticas de crítica à evolução das espécies embasando-se na bíblia

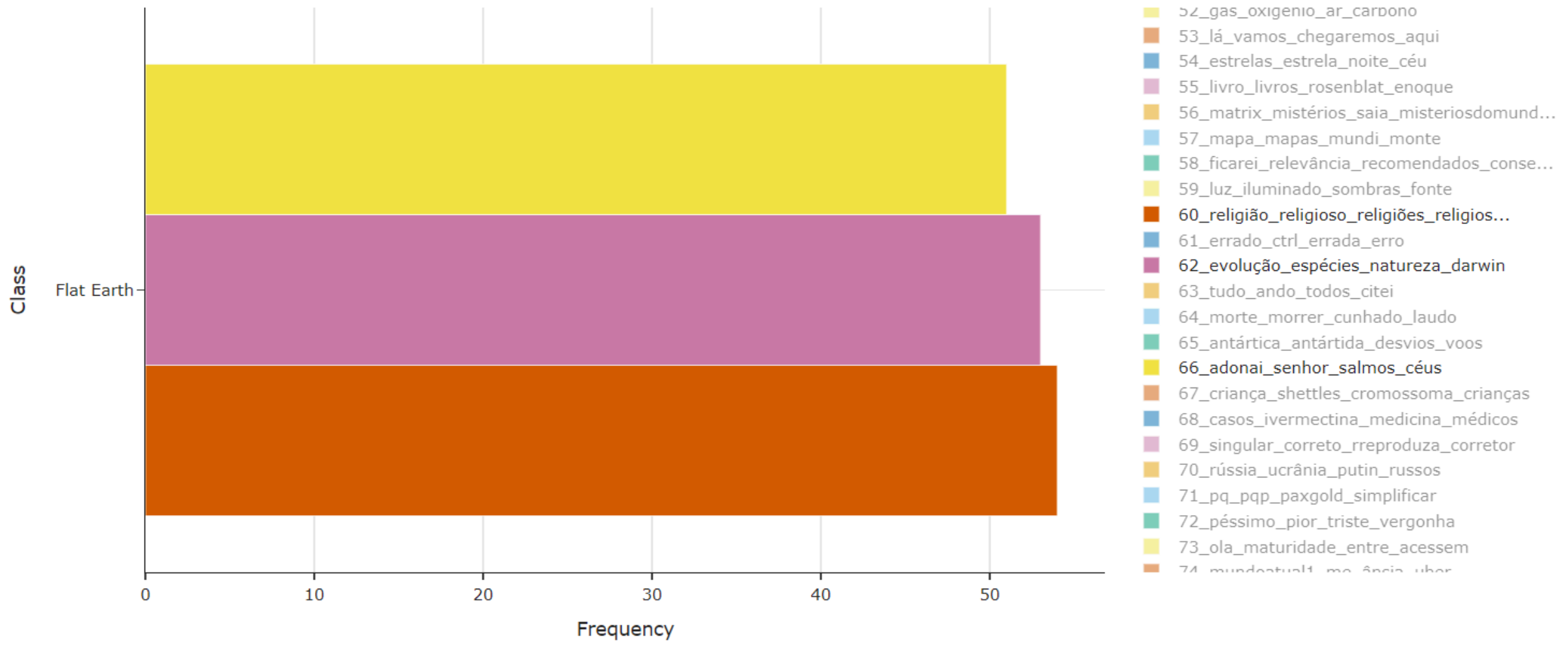

Fonte: Elaboração própria (2024).

A Figura 09 destaca a crítica à teoria da evolução das espécies, sustentada por argumentos bíblicos dentro das comunidades terraplanistas. Tópicos como "evolução", "espécies", e "Darwin" são abordados de forma a negar a validade científica da evolução, utilizando-se de interpretações religiosas para sustentar essa rejeição. Essa sobreposição de temas reflete um esforço em desacreditar teorias científicas amplamente aceitas, reforçando uma visão de mundo que coloca a narrativa bíblica como a única verdade inquestionável, opondo-se diretamente às evidências científicas.

**Figura 10.** Temáticas de radiação solar e explosões solares

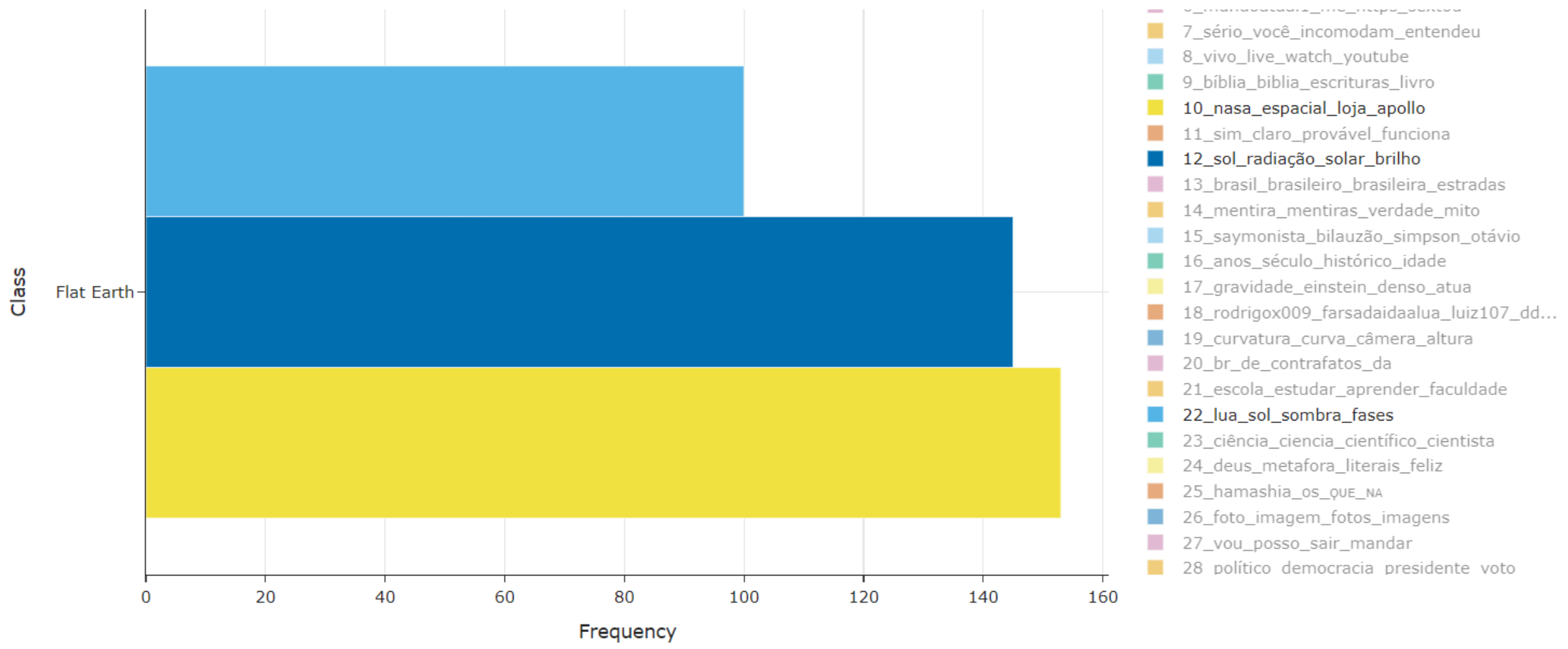

Fonte: Elaboração própria (2024).

Na Figura 10, a temática da radiação solar e explosões solares é explorada em conexão com o terraplanismo. Tópicos como "sol", "radiação", e "explosões solares" são discutidos de forma a sugerir que eventos astronômicos são manipulados ou mal interpretados pela ciência tradicional. Essa abordagem serve para reforçar a desconfiança em relação à ciência espacial e cosmológica, sustentando a ideia de que a Terra plana é a verdadeira representação da realidade, enquanto fenômenos como a radiação solar são vistos como parte de uma farsa.

**Figura 11.** Temáticas de apocalipse bíblico e pseudofrequências

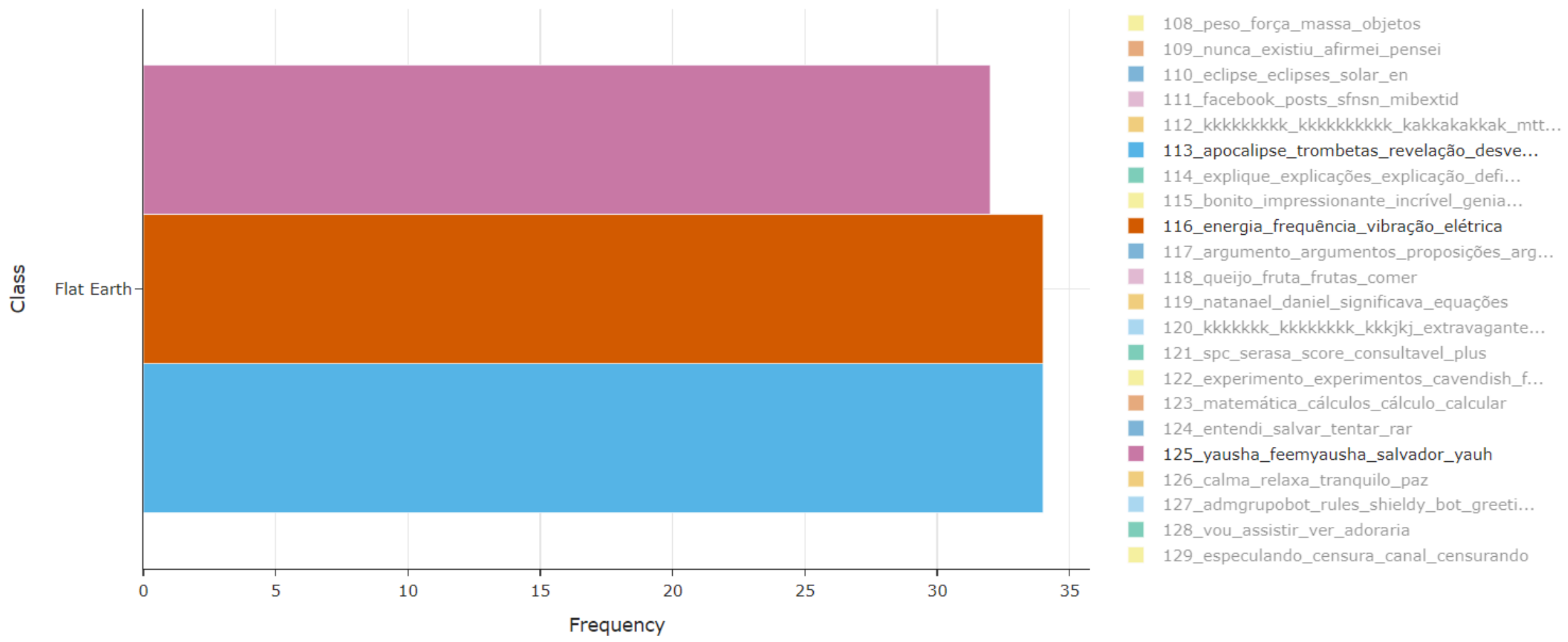

Fonte: Elaboração própria (2024).

A Figura 11 explora a interseção entre o terraplanismo e narrativas de apocalipse bíblico, juntamente com discussões sobre pseudofrequências. Tópicos como "apocalipse", "trombetas", e "vibração" mostram que essas comunidades frequentemente conectam suas crenças na Terra plana com previsões apocalípticas baseadas na bíblia. Essa interseção de temas sugere que o terraplanismo é visto não apenas como uma crença sobre a forma da Terra, mas como parte de uma visão mais ampla de um fim iminente, onde eventos cósmicos são interpretados como sinais divinos.

**Figura 12.** Temáticas de inversão de polo magnético e negação cartográfica

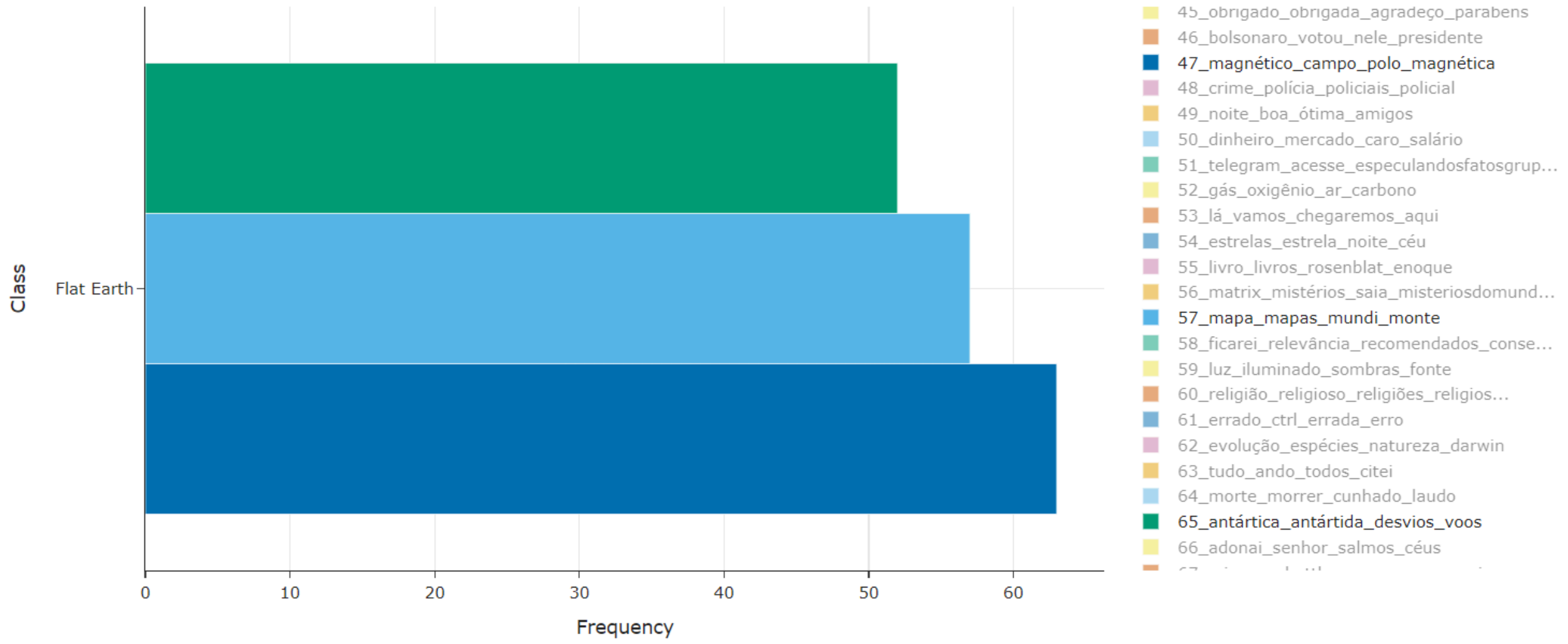

Fonte: Elaboração própria (2024).

Na Figura 12, são discutidas temáticas relacionadas à inversão do polo magnético e à negação cartográfica dentro do contexto do terraplanismo. Tópicos como "polo magnético", "mapas" e "cartografia" são abordados para questionar as interpretações científicas convencionais sobre a geografia e os fenômenos magnéticos da Terra. Essa negação dos princípios cartográficos e geofísicos reflete a tentativa dessas comunidades de construir uma nova narrativa sobre a realidade física da Terra, desacreditando as bases da geografia moderna e sustentando a crença em uma Terra plana.

## 4. Reflexões e trabalhos futuros

Para responder a pergunta de pesquisa "**como são caracterizadas e articuladas as comunidades de teorias da conspiração brasileiras sobre temáticas de terraplanismo no Telegram?**", este estudo adotou técnicas espelhadas em uma série de sete publicações que buscam caracterizar e descrever o fenômeno das teorias da conspiração no Telegram, adotando o Brasil como estudo de caso. Após meses de investigação, foi possível extrair um total de 30 comunidades de teorias da conspiração brasileiras no Telegram sobre temáticas de terraplanismo, estas somando 632.901de conteúdos publicados entre junho de 2019 (primeiras publicações) até agosto de 2024 (realização deste estudo), com 38.520 usuários somados dentre as comunidades.

Foram adotadas quatro abordagens principais: **(i)** Rede, que envolveu a criação de um algoritmo para mapear as conexões entre as comunidades por meio de convites circulados entre grupos e canais; **(ii)** Séries temporais, que utilizou bibliotecas como "Pandas" (McKinney, 2010) e "Plotly" (Plotly Technologies Inc., 2015) para analisar a evolução das publicações e engajamentos ao longo do tempo; **(iii)** Análise de conteúdo, sendo aplicadas técnicas de análise textual para identificar padrões e frequências de palavras nas comunidades ao longo dos semestres; e **(iv)** Sobreposição de agenda temática, que utilizou o modelo BERTopic (Grootendorst, 2020) para agrupar e interpretar grandes volumes de textos, gerando tópicos coerentes a partir das publicações analisadas. A seguir, as principais reflexões são detalhadas, sendo seguidas por sugestões para trabalhos futuros.

### 4.1. Principais reflexões

**Aumento expressivo das discussões sobre terraplanismo durante a Pandemia da COVID-19:** Durante o auge da Pandemia, as discussões em comunidades de terraplanismo no Telegram brasileiro aumentaram significativamente, com um crescimento de 400% no volume de menções a essa teoria entre 2019 e 2020. Esse aumento coincidiu com a disseminação de desinformação relacionada à Pandemia, sugerindo que o terraplanismo foi amplamente impulsionado por um ambiente de desconfiança generalizada nas instituições científicas e governamentais;

**Terraplanismo como porta de entrada para outras teorias conspiratórias:** As comunidades de terraplanismo não só atraem indivíduos para suas fileiras, mas também atuam como portais para que seus membros sejam expostos a outras teorias conspiratórias, como a Nova Ordem Mundial e o Ocultismo. Com um total de 1.934 links provenientes da NOM e 797 de Apocalipse e Sobrevivência, fica claro que essas teorias estão fortemente interligadas, facilitando uma transição ideológica entre diferentes narrativas desinformativas;

**A centralização das discussões em poucas comunidades:** A rede de comunidades de terraplanismo é altamente centralizada, com um pequeno número de grupos atuando como *hubs* principais para a disseminação de conteúdo. Essas comunidades centrais possuem uma capacidade desproporcional de influenciar o discurso, servindo como epicentros para a perpetuação e amplificação das narrativas terraplanistas, enquanto grupos menores orbitam;

**Interconectividade temática com a Nova Ordem Mundial:** As discussões sobre a Nova Ordem Mundial estão entre as mais frequentemente mencionadas nas comunidades de terraplanismo, com 724 links identificados como direcionados a essa temática. Isso sugere que a narrativa de um suposto controle global é frequentemente utilizada para legitimar e reforçar as crenças terraplanistas, criando sobreposição temática que fortalece as narrativas;

**Papel das mídias digitais na propagação do terraplanismo:** Termos como “canal” e “Telegram” foram identificados como centrais nas discussões, indicando que a disseminação do terraplanismo depende fortemente de plataformas digitais, guiando as mobilizações. A prevalência desses termos ao longo dos anos ressalta a importância do Telegram e outras plataformas na manutenção e expansão das comunidades terraplanistas;

**Relação entre terraplanismo e narrativas religiosas:** A análise de conteúdo revelou uma forte presença de termos religiosos, como “Deus” e “bíblia”, nas discussões sobre terraplanismo. Isso indica que o terraplanismo é frequentemente moldado e sustentado por uma visão de mundo que combina elementos religiosos com desconfiança nas ciências convencionais, reforçando a crença em uma narrativa alternativa fundamentada em interpretações literais de textos religiosos;

**A rejeição à ciência como parte central da narrativa:** Comunidades de terraplanismo frequentemente utilizam discussões sobre gravidade e evolução das espécies para rejeitar conceitos científicos estabelecidos. Essa rejeição não só questiona a forma da Terra, mas também desafia toda a estrutura do conhecimento científico, reforçando uma visão de mundo onde a ciência é vista como uma construção falaciosa destinada a enganar;

**Atração por teorias de catástrofes globais e apocalípticas:** Temas como apocalipse bíblico e inversão de pólos magnéticos estão intimamente conectados ao terraplanismo, sugerindo que essas comunidades frequentemente associam a crença na Terra plana com previsões catastróficas e com narrativas dogmáticas oriundas da espiritualidade. Essas narrativas apocalípticas funcionam como uma extensão lógica da desconfiança geral na ciência e nas autoridades, oferecendo uma explicação alternativa para eventos globais;

**Persistência e resiliência das narrativas terraplanistas:** Mesmo com os esforços de desmantelamento dessas crenças, as narrativas terraplanistas mostraram uma resiliência notável, mantendo discussões ativas e coesas ao longo dos anos. A análise de séries temporais mostra que, embora tenha havido picos de interesse, as discussões se estabilizaram em um nível ainda elevado, indicando que essas crenças se enraizaram profundamente no digital;

**Terraplanismo como um teste de comprometimento ideológico:** A interconectividade entre terraplanismo e outras teorias conspiratórias sugere que o envolvimento com essa narrativa funciona como um teste de comprometimento ideológico para seus adeptos. Uma vez que indivíduos se envolvem com o terraplanismo, eles são frequentemente expostos a uma gama mais ampla de narrativas conspiratórias, ampliando seu engajamento dentro do ecossistema conspiratório e dificultando a desconstrução de crenças.

### 4.2. Trabalhos futuros

Baseando-se nos principais achados deste estudo, várias direções promissoras podem ser exploradas para pesquisas futuras. A interseção entre o terraplanismo e outras narrativas apocalípticas sugere um campo fértil para investigações que busquem entender como essas teorias conspiratórias se conectam e se reforçam mutuamente. Por exemplo, estudos futuros poderiam focar em como as narrativas de Nova Ordem Mundial e Apocalipse e Sobrevivência funcionam como pontes para o terraplanismo, alimentando a adesão a outras crenças extremas. Além disso, seria relevante investigar como crises globais, como a Pandemia da COVID-19, catalisam a disseminação dessas teorias, oferecendo novas perspectivas sobre a evolução e a adaptação do terraplanismo em períodos de incerteza.

Outro ponto crítico para futuras investigações é a centralização observada nas redes de terraplanismo, onde um pequeno número de comunidades atua como *hubs* de disseminação. Pesquisas poderiam se concentrar em mapear essas estruturas centrais para identificar como elas mantêm a coesão e a influência dentro da rede. Compreender o papel desses *hubs* poderia fornecer insights valiosos para desenvolver estratégias de intervenção que visem interromper a disseminação de desinformação de forma mais eficaz.

A relação entre o terraplanismo e a religião também abre um leque de possibilidades para estudos futuros. Considerando que essas comunidades utilizam interpretações bíblicas para legitimar a rejeição à ciência, é fundamental explorar como essas narrativas se sustentam e se perpetuam. Investigações que abordem a psicologia por trás da fusão de crenças religiosas e teorias conspiratórias poderiam revelar as razões pelas quais essas ideias são tão resistentes a intervenções externas, facilitando o desenvolvimento de campanhas de correção mais direcionadas e eficazes.

Além disso, a importância das plataformas digitais na propagação do terraplanismo não pode ser subestimada. Estudos futuros poderiam investigar como a arquitetura dessas plataformas facilita a disseminação de desinformação e como as comunidades de terraplanismo adaptam suas estratégias de comunicação diante de restrições ou censuras em outras mídias. A análise da migração de discussões para plataformas alternativas e o impacto disso na disseminação de ideias radicais também seria uma área promissora para exploração.

Por fim, a persistência das narrativas conspiratórias ao longo do tempo, mesmo após repetidos desmentidos científicos, indica a necessidade de entender os mecanismos que permitem que essas crenças sejam reintroduzidas e ganhem tração novamente. Pesquisas poderiam focar em como esses ciclos de resiliência e reintrodução ocorrem, particularmente durante momentos críticos, e como eles podem ser interrompidos de maneira eficaz. Compreender esses processos pode ser a chave para o desenvolvimento de estratégias de longo prazo que busquem desacelerar a propagação de desinformação e proteger o debate público das consequências nocivas dessas crenças.

## 5. Referências

## 6. Biografia do autor

**Ergon Cugler de Moraes Silva** possui mestrado em Administração Pública e Governo (FGV), MBA pós-graduação em Ciência de Dados e Análise (USP) e bacharelado em Gestão de Políticas Públicas (USP). Ele está associado ao Núcleo de Estudos da Burocracia (NEB FGV), colabora com o Observatório Interdisciplinar de Políticas Públicas (OIPP USP), com o Grupo de Estudos em Tecnologia e Inovações na Gestão Pública (GETIP USP), com o Monitor de Debate Político no Meio Digital (Monitor USP) e com o Grupo de Trabalho sobre Estratégia, Dados e Soberania do Grupo de Estudo e Pesquisa sobre Segurança Internacional do Instituto de Relações Internacionais da Universidade de Brasília (GEPSI UnB). É também pesquisador no Instituto Brasileiro de Informação em Ciência e Tecnologia (IBICT), onde trabalha para o Governo Federal em estratégias contra a desinformação. Brasília, Distrito Federal, Brasil. Site: https://ergoncugler.com/.